\documentclass{jfm}
\usepackage[version=4]{mhchem}
\usepackage{amsmath}
\usepackage{graphicx}
\usepackage{epstopdf, epsfig}
\usepackage{amsfonts}
\usepackage{amssymb}
\usepackage{float}
\usepackage{xcolor}

\usepackage{bm}
\usepackage[acronym]{glossaries}
\usepackage{soul}
\usepackage{tikz}
\usetikzlibrary{shapes,arrows}
\usetikzlibrary{positioning}
\usetikzlibrary{external}

\usepackage{dcolumn}
\usepackage{bm}
\usepackage{hyperref}
\usepackage{multirow}
\usepackage{comment}
\usepackage{xcolor}
\usepackage{tikz}
\usetikzlibrary{arrows,decorations.markings,shapes,positioning}
\usepackage{subcaption}

\usepackage{todonotes}

\usepackage{color,soul}
\usepackage[normalem]{ulem}
\usepackage{soul}

\shorttitle{Kinetic framework with consistent hydrodynamics for shallow water equations}
\shortauthor{S .A. Hosseini and I. V. Karlin}

\title{Kinetic framework with consistent hydrodynamics for  shallow water equations}

\author{S. A. Hosseini\aff{1\corresp{\email{}shosseini@ethz.ch}},
 \and I. V. Karlin\aff{{1\corresp{\email{}ikarlin@ethz.ch}}}
}
\affiliation{
\aff{1}Department of Mechanical and Process Engineering, ETH Zurich, 8092 Zurich, Switzerland.}

\begin{document}
\maketitle
\begin{abstract}
We present a novel discrete velocity kinetic framework to consistently recover the viscous shallow water equations. The proposed model has the following fundamental advantages and novelties: (a) A novel interpretation and general framework to introduce forces, (b) the possibility to consistently split pressure contributions between equilibrium and a force-like contribution, (c) consistent recovery of the viscous shallow water equations with no errors in the dissipation rates, (d) independent control over bulk viscosity, and (e) consistent second-order implementation of forces. As shown through a variety of different test cases, these features make for an accurate and stable solution method for the shallow-water equations.
{
\centering
    \hspace{4cm}\includegraphics[width=0.4\textwidth]{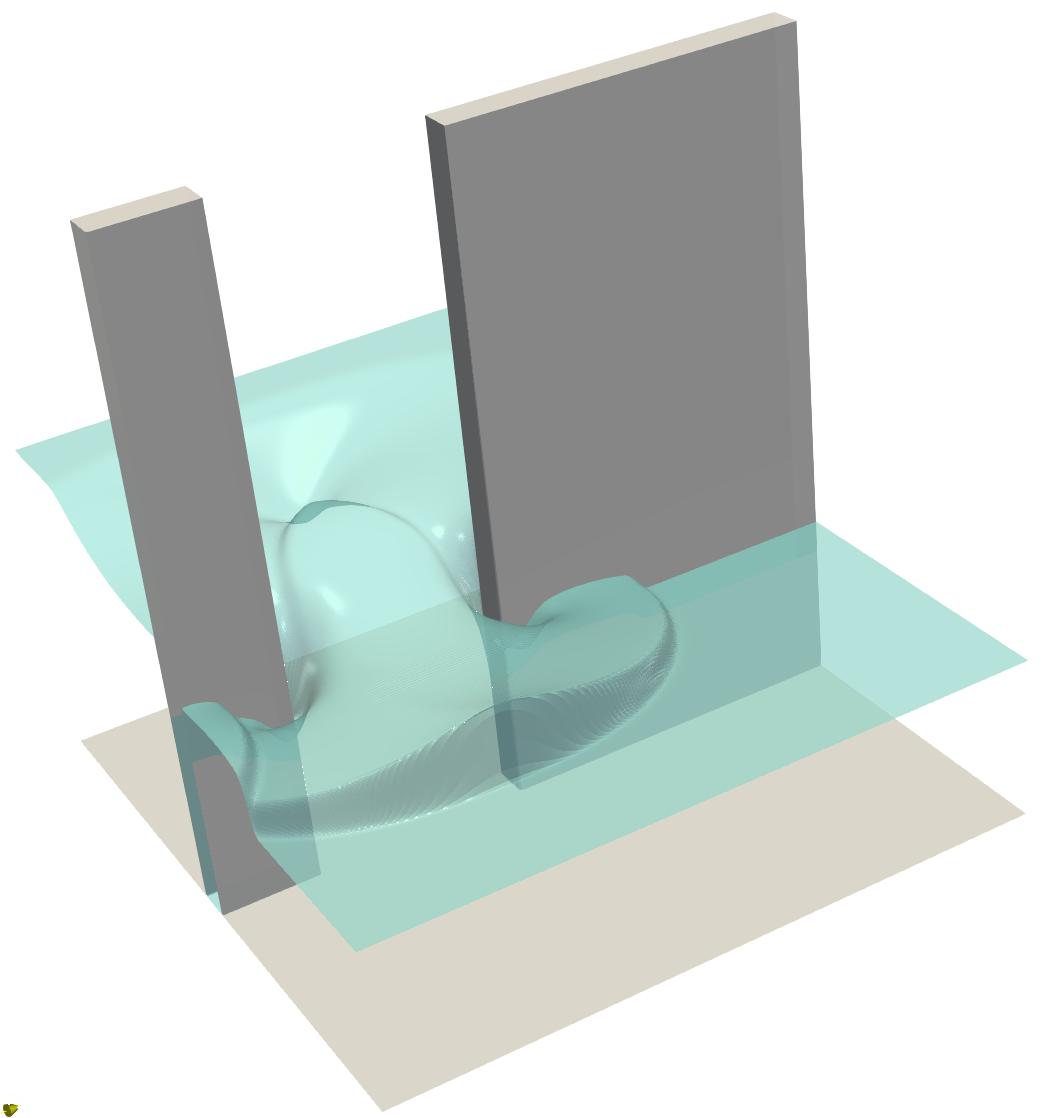}
}  
\end{abstract}


\section{\label{sec:introduction}Introduction}
The shallow water equations (SWEs) constitute a system of equations that govern the evolution of fluid flow in situations where the horizontal length scales greatly exceed the vertical depth of the fluid layer~\citep{tan1992shallow,vreugdenhil2013numerical}. Derived from the depth-averaged incompressible Navier–Stokes (NS) equations under the hydrostatic and Boussinesq approximations, SWEs are extensively used in geophysical fluid dynamics to model atmospheric and oceanic phenomena, including tsunamis, storm surges, tidal flows, and large-scale atmospheric motions \citep{vallis2017atmospheric, leVeque2002finite, bresch2009shallow}.

Mathematically, SWEs represent conservation laws for mass and horizontal momentum, typically expressed in conservative or non-conservative form, with source terms accounting for variable bottom topography, Coriolis forcing, and, in some formulations, bed friction and other dissipative processes \citep{ghil1981numerical}. In one or two spatial dimensions, the equations take the form of a nonlinear hyperbolic system that admits wave-like solutions and can exhibit discontinuities, such as shocks or hydraulic jumps, even in the absence of explicit viscosity \citep{whitham1974linear}.

Due to their non-linearity and the need to accurately resolve features such as moving wet/dry interfaces and complex boundary geometries, numerical solution of SWEs remains a central topic in computational fluid dynamics~\citep{toro2024computational}. A wide range of discretization techniques has been developed, including finite difference, finite volume, and discontinuous Galerkin methods, each with trade-offs in terms of stability, conservation, and computational efficiency \citep{leveque1998balancing, cockburn2000development}.

With the growing success of the lattice Boltzmann method (LBM) in modeling flows in the limit of incompressibility, it was extended to a variety of more complex flows such as compressible~\citep{frapolli2015entropic,wilde2020semi,farag2020pressure,hosseini2024lattice}, non-ideal and multi-phase~\citep{chen2014critical,hosseini2023lattice}, non-Newtonian etc. Given that SWE's bear a lot of similarities with the NS equations, efforts to devise lattice Boltzmann models for them started as early as in 1999. To the author's knowledge, the first attempt at proposing a lattice Boltzmann model to solve the shallow water equations was documented in \citep{salmon1999lattice} where the author proposed a modified form of the discrete equilibrium distribution function to properly recover the barotropic equation of state in the shallow water equations.
Subsequently, starting in the early 2000's a large number of articles started considering this topic and proposed improvements and extensions to Salmon's solver such as bed topography, wind~\citep{linhao2005wind}, Coriolis force effects~\citep{dellar2005shallow}, introduction of turbulence models~\citep{linhao2005wind,xin2010lattice}, handling of wet-dry front evolution~\citep{liu2014lattice,liu2016second} and fluid-solid interaction~\citep{geveler2010lattice,de2014lattice}. To date, to the best of the author's knowledge, Salmon's equilibrium remains the most widely used solution proposed in the literature for the shallow-water equations as majority of subsequent publications have used this equilibrium-- see, for instance, \citep{venturi2020new,de2023comparison}. As noted in \citep{dellar2002nonhydrodynamic}, stability has always been an issue of concern in the development of LBM for SWE's and one of the main reasons Salmon's equilibrium has been preferred over a traditional second-order polynomial equilibrium. Widespread efforts during the past decades in improving stability have manifested in the form of variants of the model with multiple-relaxation-time realizations; see, for instance,~\citep{venturi2020new,venturi2021modelling,venturi2020forcing,de2023comparison,de2017central}. While the use of mechanisms such as individual relaxation rates for ghost moments can affect stability properties through forms of hyperviscosity, one major issue is that both the second-order Hermit expansion and Salmon's modified equilibrium are subject to Galilean-variance issues in both shear and bulk viscosity as the cut-off in the expansion at order two leads to incorrect third-order equilibrium moments; see discussion on that topic for instance in \citep{li2012coupling,hosseini2020compressibility,hosseini2022towards,hosseini2023lattice,prasianakis2007lattice,prasianakis2009lattice}. This issue, demonstrated and derived in the Appendix \ref{app:on_diff}, points to another big gap in the lattice Boltzmann literature for SWEs: all models target the inviscid SWEs and there is little to none that has been done for the viscous SWEs.
Coming back to stability issues, this is an important point to note, as NS level instabilities can only be overcome with proper discrete equilibria. A proof of this statement is our previous studies on the concept of asymptotic freedom~\citep{hosseini2024asymptotic,hosseini2025linear}.Furthermore, while errors in dissipation rates for the second-order polynomial family of discrete equilibria are usually cited as scaling with the cube of the Mach number~\citep{kruger2017lattice}, this is only valid for the isothermal ideal equation of state lattice Boltzmann solvers. For nonideal fluids, the ratio of pressure to density deviates from the lattice reference temperature and leads to errors scaling linearly with the Mach number~\citep{hosseini2022towards}. To sum up: (a) These errors preclude all of these solvers from properly capturing NS level dynamics and can accumulate to be significant. (b) While Euler level dynamics is not \emph{directly} affected in terms of dispersion rate in the hydrodynamic limit, they are still affected through the spectral dissipation rate, which dictates the stability of the discrete solver. Stabilizing effects can, for example, be seen in the context of compressible flows, in \citep{hosseini2020compressibility,renard2021improved}. 
In contrast to compressible fluids where dissipation properties have been probed both theoretically and numerically, and proper corrections have been devised to restore correct hydrodynamics in the limit of vanishing wave numbers, the shallow-water literature lacks detailed studies and discussions.

In the present work, we address the issues of fundamental importance listed above: We propose a family of discrete kinetic models with a novel interpretation of the force term as a relaxation term, consistently recovering the \emph{full viscous shallow water equations} in the hydrodynamic limit with \emph{independent control over bulk viscosity}, which to the authors' knowledge is the first proposal of its kind. 
In addition, we present a consistent Lagrangian discretization in space and time that maintains second-order accuracy with external body force, as this is a topic of importance in LBM \citep{guo2002discrete,peng2017second,li2019evaluation,venturi2020forcing} This allows us to derive a second-order accurate formulation with forcing. The proposed family of solvers is then applied to different benchmark cases and shown to be stable and correctly recover reference solutions.

\section{Consistent lattice Boltzmann model for shallow water equations\label{sec:numericalMethod}}
\subsection{Target shallow water equations}
We begin with a brief overview of the basic shallow water system, see textbooks \citep{tan1992shallow} for in-depth discussion. The $D$-dimensional SWEs, with $D\leq 2$ can be derived from the $(D+1)$-dimensional incompressible Navier--Stokes equations with a free moving surface boundary condition.
The resulting system is that of conservation of a liquid column height $h$ and the corresponding momentum $h\bm{u}$ written as,
    \begin{align}
       & \partial_t h + \bm{\nabla}\cdot h\bm{u} = 0,
    	\label{eq:shallow_water_h}
    	\\
    	&\partial_t h\bm{u} + \bm{\nabla}\cdot h\bm{u}\otimes\bm{u} + \bm{\nabla}P + \bm{\nabla}\cdot T_{\rm NS}= \bm{F}_{\rm ext}.
    	\label{eq:shallow_water}
    \end{align}
{Here $P$ is the nonlinear barotropic pressure,
\begin{equation}
    \label{eq:P_SWE}
    P=\frac{g}{2}h^2,
\end{equation}
with $g$ the gravitational acceleration and $\bm{u}$ the depth-averaged fluid velocity. Furthermore, 
$T_{\rm NS}$ in \eqref{eq:shallow_water} represents the viscous stress tensor. While different forms have been employed to model the viscous stress, the Navier--Stokes (NS) viscous stress remains the preferred option \citep{rodriguez2005navier,bresch2011mathematical,bresch2003existence},
\begin{equation}\label{eq:TNS}
    T_{\rm NS} = {-}h\nu\left(\bm{\nabla}\bm{u} + \bm{\nabla}\bm{u}^\dagger - \frac{2}{D}(\bm{\nabla}\cdot\bm{u})\bm{I}\right) {-} h\eta (\bm{\nabla}\cdot\bm{u})\bm{I},
\end{equation}
with $\bm{I}$ the unit matrix.}
Note that, $h[{\rm m}]$ represents the height of the water column in meters, $g$ is the gravitational acceleration in $[{m}/{s}^2]$, $\nu$ and $\eta$ are the kinematic and the bulk viscosity, respectively, in $[{\rm m}^2/{\rm s}]$ while the pressure $P$ \eqref{eq:P_SWE} has units of {$[{\rm m}^3/{\rm s}^2]$}. 

\begin{figure}
	\centering
		\includegraphics[width=0.4\linewidth]{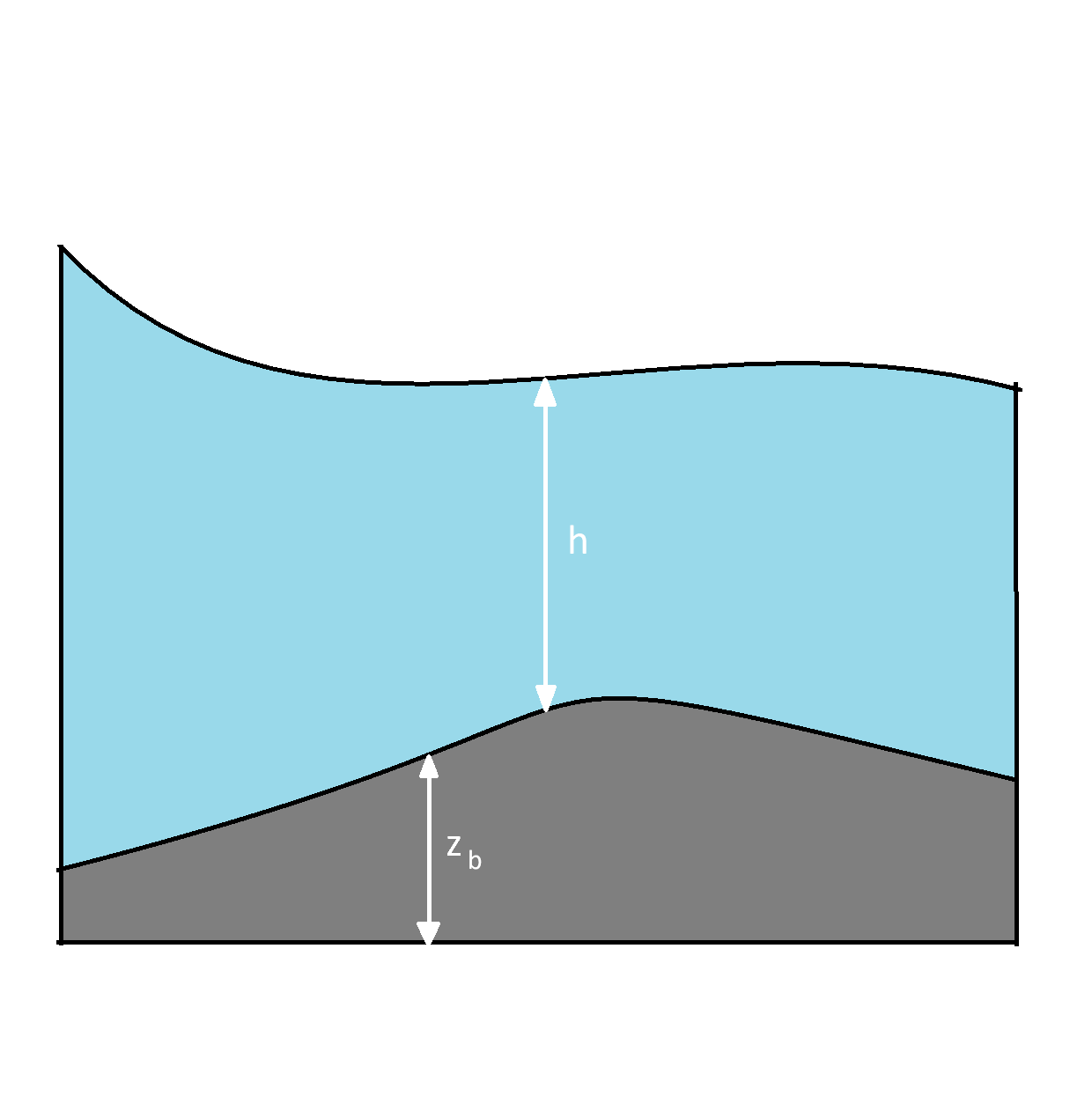}
	\caption{Schematics of shallow water equations variables.}
	\label{image_swe}
\end{figure}

Finally, the {external} force term $\bm{F}_{\rm ext}$ can be used to account for a variety of effects, for instance,
\begin{equation}
    \bm{F}_{\rm ext} =  \bm{F}_{\rm b} + \bm{F}_{\rm sb} + \bm{F}_{\rm sw}+\bm{F}_{\rm c}.
\end{equation}
Here $\bm{F}_{\rm b}$ accounts for the bed topography and is defined as,
\begin{equation}
    \bm{F}_{\rm b} = -gh\bm{\nabla} z_{\rm b},
\end{equation}
with $z_{\rm b}$ the bed height, see schematics in Fig. \ref{image_swe} for an explanation.
Furthermore, $\bm{F}_{\rm sb}$ and $\bm{F}_{\rm sw}$ account for the viscous stress at the bed and at the water column surface in contact with air, respectively. The bed shear stress is usually given by the depth-averaged velocities as,
\begin{equation}
    F_{{\rm sb,} {\alpha}} = -C_{\rm b} u_{\alpha} \lvert \bm{u}\lvert,
\end{equation}
with,
\begin{equation}
    C_{\rm b} = \frac{g n_{\rm b}^2}{h^{1/3}},
\end{equation}
where $n_{\rm b}$ is known as Manning's coefficient. The wind stress is also usually computed as,
\begin{equation}
    F_{\rm sw, {\alpha}} = \frac{\rho_{\rm air}}{\rho_{\rm water}}C_{\rm w} u_{{\rm w,}{\alpha}} \lvert \bm{u}_{\rm w}\lvert,
\end{equation}
where $\rho_{\rm air}$ is the density of air,  $\bm{u}_{\rm w}$ is the wind velocity and $C_{\rm w}$ the resistance coefficient. 
Finally, the Coriolis force $\bm{F}_{\rm c}$ is defined as,
\begin{equation}\label{eq:coriolis}
    \bm{F}_{\rm c} = h f_{\rm c}\begin{bmatrix}
u_y\\-u_x        
    \end{bmatrix},
\end{equation}
where $f_{\rm c}$ is the Coriolis parameter. 

{Shallow water equations \eqref{eq:shallow_water_h} and \eqref{eq:shallow_water} are the target macroscopic equations.
In the following, we shall present a kinetic model that recovers the shallow-water system in the hydrodynamic limit, together with its lattice Boltzmann realization.}

\subsection{Discrete velocity kinetic model}

{We consider the standard discrete velocity set {$\bm{v}_i=c\bm{c}_i$} in two dimensions ($D=2$) with nine velocities ($Q=9$),
\begin{equation}
    \label{eq:D2Q9c}
    \bm{c}_i=(c_{ix},c_{iy}),\ c_{i\alpha}\in\{-1,0,1\},
\end{equation}
and denote as $f_i(\bm{x},t)$ the corresponding populations of the D$2$Q$9$ lattice \eqref{eq:D2Q9c}. 
The D$2$Q$9$ lattice is characterized with the lattice speed of sound,
\begin{equation}
    \label{eq:cs}
    \varsigma=\frac{1}{\sqrt{3}}c.
\end{equation}
{Below, we consider lattice units by setting $c=1$.}

Motivated by the generic approach initialized in \citep{sawant_consistent_2021,sawant_consistent_2022,hosseini2022towards}, we introduce a kinetic equation,}
{\begin{equation}\label{eq:final_kinetic_model}
	\partial_t f_i + \bm{c}_i\cdot\bm{\nabla} f_i = -\frac{1}{\tau}\left(f_i-f_i^{\rm eq}(h,\bm{u})\right) 	+ \frac{1}{\lambda}(f_i^*(h,\bm{u};\lambda)-f_i^{\rm eq}(h,\bm{u})).
\end{equation}
}
{The first term in the right-hand side represents the lattice Bhatnagar--Gross--Krook (LBGK) relaxation to the local equilibrium $f_i^{\rm eq}$ with a \emph{bare} relaxation time $\tau$. The second term is a generalized forcing, characterized by \emph{shifted-equilibrium} populations $f_i^*$ and a time parameter $\lambda$ which will be specified below.
The locally conserved fields, the water column height $h$ and the flow velocity $\bm{u}$, are defined via the zeroth- and the first-order moments of the populations, respectively,
\begin{align}
    &h=\sum_{i=1}^Q f_i,\label{eq:def_h}\\
    &h\bm{u}=\sum_{i=1}^Q \bm{c}_if_i.\label{eq:def_u}
\end{align}
}

{We proceed with defining the equilibrium and the shifted-equilibrium populations. To that end, we first introduce a triplet of functions in two variables, $\xi_{\alpha}$ and $\zeta_{\alpha\alpha}$,
    \begin{equation}\label{eq:phi}
    \begin{split}
    	&	\Psi_{0}(\xi_{\alpha},\zeta_{\alpha\alpha}) = 1 - \zeta_{\alpha\alpha}, 
    	\\
    	&	\Psi_{1}(\xi_{\alpha},\zeta_{\alpha\alpha}) = \frac{\xi_{\alpha} + \zeta_{\alpha\alpha}}{2},
    	\\
    	&	\Psi_{-1}(\xi_{\alpha},\zeta_{\alpha\alpha}) = \frac{-\xi_{\alpha} + \zeta_{\alpha\alpha}}{2}.
        \end{split}
        \end{equation}
In order to save notation, we introduce the vectors of the parameters, 
\begin{align}
    &\bm{\xi}=(\xi_x,\xi_y)\\
    &\bm{\zeta}=(\zeta_{xx},\zeta_{yy}),
\end{align}
and consider a product-form,
\begin{equation}\label{eq:PF}
    \Theta_i(\bm{\xi},\bm{\chi})=\Psi_{c_{ix}}(\xi_x,\chi_{xx})\Psi_{c_{iy}}(\xi_y,\chi_{yy}),\ i=1,\dots Q.
\end{equation}
Note that, because the components of the discrete velocities $\bm{c}_i$ take the values $\pm1$ or $0$, cf. \eqref{eq:D2Q9c}, the elements of the product-form \eqref{eq:PF} are uniquely defined for any $i=1,\dots,9$.}

In order to define the equilibrium, we set the parameters as follows: 
    \begin{align}
    &\xi_{\alpha}^{\rm eq}=u_{\alpha},\label{eq:def_xi_eq}\\
    &\zeta_{\alpha\alpha}^{\rm eq}=\frac{P_0}{h}+u_{\alpha}^2.\label{eq:def_zeta_eq}
    \end{align}
{Here we have introduced a \emph{reference pressure} $P_0(h)$ \citep{hosseini2022towards} as a yet unspecified parameter which we shall discuss below.}
{With the definitions \eqref{eq:def_xi_eq} and \eqref{eq:def_zeta_eq} in the functions \eqref{eq:phi}, 
the local equilibrium populations are represented with a product-form,}
    \begin{equation}\label{eq:LBMeq}
         f_i^{\rm eq}(h,\bm{u})= h\Psi_{c_{ix}}\left(u_x,\frac{P_0}{h}+u_{x}^2\right)\Psi_{c_{iy}}\left(u_y,\frac{P_0}{h}+u_{y}^2\right).
    \end{equation}

{Moving onto the shifted-equilibrium populations $f_i^*$, the parameters $\xi_\alpha$ and $\zeta_{\alpha\alpha}$ in the triplet of functions  \eqref{eq:phi} are set as follows:
\begin{align}
	&\xi_{\alpha}^{*} = u_{\alpha}+{\lambda}\frac{F_{\alpha}}{h},\label{eq:xistar}	\\
	  &\zeta_{\alpha\alpha}^{*} = \left(\frac{P_0}{h} +u_{\alpha}^2\right) + {\lambda}\frac{\Phi_{\alpha\alpha}}{h}.\label{eq:zetastar}
\end{align}
When compared to their equilibrium counterparts, the added terms in \eqref{eq:xistar} and \eqref{eq:zetastar} are, respectively,
\begin{align}
&    {F}_\alpha = -\partial_\alpha(P- P_0) + {F}_{{\rm ext},\alpha},\label{eq:u_shift}\\
\label{eq:correction}
&	   \Phi_{\alpha\alpha} = \partial_{\alpha}\left(h u_{\alpha} \left(u_{\alpha}^2 + \frac{3P_{0}}{h}-3\varsigma^2\right)\right) + \left(\frac{\eta}{\tau} - P_0\left(\frac{D+2}{D}-\frac{\partial \ln P_0}{\partial \ln h}\right)\right) (\bm{\nabla}\cdot\bm{u}).
\end{align}
We shall interpret and discuss the latter terms below. 
For now, we finalize the construction of the {shifted-equilibrium} populations by using again the triplet of functions \eqref{eq:phi} but with the parameters set according to \eqref{eq:xistar} and \eqref{eq:zetastar}. 
This leads to the product-form, similar to \eqref{eq:LBMeq},}
\begin{equation}\label{eq:LBMstar}
     f_i^*(h,\bm{u};{\lambda})=h\Psi_{c_{ix}}\left(u_x+{\lambda}\frac{F_{x}}{h},\frac{P_0}{h}+u_{x}^2+{\lambda}\frac{\Phi_{xx}}{h}\right)
     \Psi_{c_{iy}}\left(u_y+{\lambda}\frac{F_{y}}{h},\frac{P_0}{h}+u_{y}^2+{\lambda}\frac{\Phi_{yy}}{h}\right),
\end{equation}
which allows us to define the forcing term in the kinetic equation \eqref{eq:final_kinetic_model}.
{Thus, the kinetic model \eqref{eq:final_kinetic_model} is fully defined with the input of shallow water pressure $P$ \eqref{eq:P_SWE} and the reference pressure $P_0$. Comments are in order:
\begin{itemize}
\item The role of the reference pressure $P_0$, featured in \eqref{eq:def_zeta_eq} and in \eqref{eq:u_shift} is to split the pressure $P$ \eqref{eq:P_SWE} into two contributions: The reference pressure is provided via the equilibrium while the rest, $P-P_0$, is due to the forcing term \eqref{eq:u_shift} featured in the shifted-equilibrium \eqref{eq:LBMstar}. 
The two distinguished cases of the splitting are:
\begin{align}
    &{\rm Case\ A:}\ P_0=\varsigma^2 h,\label{eq:A}\\
    &{\rm Case\ B:}\ P_0=P.\label{eq:B}
\end{align}
In the Case A \eqref{eq:A}, the reference $P_0$ is the standard lattice Boltzmann pressure, defined via the lattice speed of sound \eqref{eq:cs}. With the forcing term neglected, the kinetic equation \eqref{eq:final_kinetic_model} becomes the standard lattice Boltzmann kinetic equation for athermal fluid. 
Case B \eqref{eq:B} can be termed the minimal forcing, as the pressure $P$ is provided by the equilibrium, while the corresponding contribution from the forcing vanishes in \eqref{eq:u_shift}. Below, we shall compare the actual numerical performance of both the Cases A and B.

\item The function \eqref{eq:correction} is a correction term that addresses two issues.
First, we remind that the discrete velocity bias of the D2Q9 lattice, $c_{i\alpha}^3=c_{i\alpha}$, restricts Galilean invariance of the kinetic model. Thus, the first term in \eqref{eq:correction} restores Galilean invariance in the hydrodynamic limit, along the line of earlier proposals \citep{prasianakis2007lattice,saadat_2021_extended,hosseini2022towards}. 
The second term in \eqref{eq:correction} allows for achieving independent bulk viscosity $\eta$ in the hydrodynamic limit. 
Similar strategies to introduce bulk viscosity for ideal gas equation of state, both pure substances and mixtures, have been proposed, e. g. in \citep{renard2021improved,sawant_consistent_2022}.
\item In order to gain some intuition about relative magnitude of the relaxation parameters in \eqref{eq:final_kinetic_model}, it is instructive to rewrite the right hand side of Eq.\ \eqref{eq:final_kinetic_model} as follows:
\begin{equation}\label{eq:final_kinetic_model2}
	\partial_t f_i + \bm{c}_i\cdot\bm{\nabla} f_i = -\frac{1}{\theta}\left(f_i-f_i^{\rm eq}(h,\bm{u})\right) 
    - \frac{1}{\lambda}(f_i-f_i^*(h,\bm{u};\lambda)),
\end{equation}
where we have introduced the \emph{effective} relaxation time to the equilibrium, 
\begin{equation}\label{eq:theta}
    \theta=\frac{\tau\lambda}{\lambda-\tau}.
\end{equation}
The form \eqref{eq:final_kinetic_model2} reveals the two relaxation processes executed in parallel: Relaxation to the equilibrium with the effective relaxation time $\theta$ \eqref{eq:theta} and the relaxation to the shifted-equilibrium with the relaxation time $\lambda$. Positivity of the effective relaxation time $\theta\ge 0$ requires a hierarchy of relaxation times to hold, 
\begin{equation}
    \label{eq:hierarchy}\lambda\ge \tau.
\end{equation} 
Moreover, the effective relaxation time $\theta$ tends to become the bare relaxation time $\tau$ when the relaxation to the equilibrium becomes dominant,
\begin{equation}
    \label{eq:dominance}
    \theta\to\tau\ {\rm for}\ \lambda\gg\tau.
\end{equation} 
\item In order to see that the last term  in \eqref{eq:final_kinetic_model} is a (generalized) forcing, let us neglect, for simplicity of presentation, the correction term \eqref{eq:correction}, leaving only the forcing term in \eqref{eq:u_shift}; then:
\begin{equation}
    \lim_{\lambda\to0}\frac{1}{\lambda}(f_i^*(h,\bm{u};\lambda)-f_i^{\rm eq}(h,\bm{u}))=\frac{\partial \Theta_i(\bm{\xi},\bm{\chi})}{\partial \bm{\xi}}\Bigg|_{\bm{\xi}^{\rm eq},\bm{\chi}^{\rm eq}}\cdot\bm{F}.
\end{equation}
This is a special form of the forcing in the model kinetic equations such as BGK--Vlasov--Enskog, cf.\ e. g. \citep{hosseini2022towards}. However, the new and more general representation of the forcing in the form of a relaxation shall be advantageous when deriving the efficient lattice Boltzmann realization below.
    \item The Chapman--Enskog analysis of the kinetic model \eqref{eq:final_kinetic_model} in Appendix \ref{ap:CE_DVBE} reveals that the target viscous shallow water equations \eqref{eq:shallow_water_h} and \eqref{eq:shallow_water} are recovered with the kinematic viscosity defined by the reference pressure,
    \begin{equation}
	\nu=\tau \left(\frac{P_0}{h}\right).\label{eq:visc_gen}
\end{equation}
\item As already mentioned, the bulk viscosity $\eta$ is maintained due to the second term in the correction \eqref{eq:correction}. We note that, neglect of this term leads to a fixed bulk viscosity defined by the reference pressure,
\begin{equation}
    \label{eq:bulk_0}
    \eta_0=\tau \left(\frac{P_0}{h}\right)\left(\frac{D+2}{D}-\frac{\partial \ln P_0}{\partial \ln h}\right).
\end{equation}
{We also note that, for the reference pressure $P_0=P$ of Case B \eqref{eq:B}, the bulk viscosity \eqref{eq:bulk_0} vanishes for $D=2$ and the quadratic dependence of the pressure \eqref{eq:P_SWE} on the water column height.}
\end{itemize}
}

{In summary, the kinetic equation \eqref{eq:final_kinetic_model} recovers the shallow water equations in the hydrodynamic limit. It is important to stress that the nonlinear pressure $P$ \eqref{eq:P_SWE} is recovered regardless of the choice of the reference pressure in the kinetic setting. 
Equally important is to note that the hydrodynamic limit does not depend on the slow relaxation time $\lambda$, which remains a free parameter at this 
step of analysis. In particular, kinematic viscosity \eqref{eq:visc_gen} depends only on the fast relaxation time $\tau$ but not on the slow time $\lambda$. In the next section, we shall proceed with the derivation of the corresponding lattice Boltzmann discrete time-space realization for shallow water hydrodynamics.}

\subsection{Derivation of the lattice Boltzmann equation}
{We follow a procedure first introduced by \cite{he_1998_novel} for LBGK and extended to kinetic models of the type \eqref{eq:final_kinetic_model} by \cite{ansumali_2007_quasi} and \cite{sawant_consistent_2021}.}
The key step in the derivation of  the lattice Boltzmann equations from the discrete velocity model \eqref{eq:final_kinetic_model} is the integration along characteristics, here the discrete velocities, over a time $\delta t$ which leads to,
\begin{equation}\label{eq:char}
    f_i(\bm{x}+\bm{c}_i\delta t, t+\delta t) - f_i(\bm{x}, t) = \int_t^{t+\delta t}\left[\frac{1}{\tau}\left(f_i^{\rm eq} - f_i\right) 
    + {\frac{1}{\lambda}}(f_i^{*} - f_i^{\rm eq})\right]dt'.
\end{equation}
The integral on the right hand side is approximated by a trapezoidal rule,
\begin{multline}\label{eq:trapezoidal}
    \int_t^{t+\delta t}\left[\frac{1}{\tau}\left(f_i^{\rm eq} - f_i\right) +{\frac{1}{\lambda}} (f_i^{*} - f_i^{\rm eq})\right]dt' = \frac{\delta t}{2\tau}\left(f_i^{\rm eq}(\bm{x},t) - f_i(\bm{x},t)\right) + \frac{\delta t}{2{\lambda}}\left(f_i^{*}(\bm{x},t) - f_i^{\rm eq}(\bm{x},t)\right) \\+ \frac{\delta t}{2\tau}\left(f_i^{\rm eq}(\bm{x}+\bm{c}_i\delta t, t+\delta t) - f_i(\bm{x}+\bm{c}_i\delta t, t+\delta t)\right) \\ + \frac{\delta t}{2{\lambda}}\left(f_i^{*}(\bm{x}+\bm{c}_i\delta t, t+\delta t) - f_i^{\rm eq}(\bm{x}+\bm{c}_i\delta t, t+\delta t)\right) + \mathcal{O}(\delta t^3).
\end{multline}
Retaining terms up to second order renders the discrete system implicit in time. 
To remove the implicitness, the following transformation of variables is introduced \citep{he_1998_novel,ansumali_2007_quasi,sawant_consistent_2021},
\begin{equation}\label{eq:transform}
    \bar{f}_i(\bm{x},t) = f_i(\bm{x},t) - \frac{\delta t}{2\tau}\left(f_i^{\rm eq}(\bm{x},t) - f_i(\bm{x},t)\right) - \frac{\delta t}{2{\lambda}}\left(f_i^{*}(\bm{x},t) - f_i^{\rm eq}(\bm{x},t)\right).
\end{equation}
With the transform \eqref{eq:transform} in \eqref{eq:char} and \eqref{eq:trapezoidal}, we are led to the following discrete set of equations for the populations $\bar{f}_i$,
\begin{equation}\label{eq:coll_stream}
 {   \bar{f}_i(\bm{x}+\bm{c}_i\delta t, t+\delta t)= \bar{f}_i(\bm{x}, t) + 2\beta\left(f_i^{\rm eq}(\bm{x},t) - \bar{f}_i(\bm{x},t)\right) + {\frac{\delta t}{\lambda}}\left(1-\beta\right) \left(f_i^{*}(\bm{x},t) - f_i^{\rm eq}(\bm{x},t)\right),}
\end{equation}
where $\beta\in[0,1]$ is the relaxation parameter,
\begin{equation}\label{eq:beta}
    {\beta=\frac{\delta t}{2\tau+\delta t}.}
\end{equation}

{We further note that, in order to close the system \eqref{eq:coll_stream}, the fields $h(f)$ and $\bm{u}(f)$ entering the local and the shifted equilibrium populations, $f^{\rm eq}_i$ and $f^*_i$, need to be evaluated using the transformed populations $\bar{f}_i$. To that end, 
we evaluate the moments of Eq.\ \eqref{eq:transform} to obtain the following relation between the zeroth- and first-order moments of the populations $f_i$ and of the transformed populations $\bar{f}_i$,
\begin{align}
    h(\bar{f})&=h(f),\label{eq:trans_h}\\
    h\bm{u}(\bar{f})&=h\bm{u}(f)-\frac{\delta t}{2\lambda}\left(h\bm{u}(f)+\lambda \bm{F}(f)-h\bm{u}(f)\right)\nonumber\\
    &=h\bm{u}(f)-\frac{\delta t}{2}\bm{F}(f).\label{eq:trans_u1}
\end{align}
Furthermore, if the force $\bm{F}$ \eqref{eq:u_shift} is the function of the water column height $h$ only, then thanks to \eqref{eq:trans_h}, we have,
\begin{equation}
    \bm{F}(\bar{f})=\bm{F}(f).
\end{equation}
With this, the transform of the flow velocity is written explicitly,
\begin{align}
    h\bm{u}({f})&=h\bm{u}(\bar{f})+\frac{\delta t}{2}\bm{F}(\bar{f}),\label{eq:u_LBM}
\end{align}
Note that elementary transformation \eqref{eq:u_LBM} is not applicable for some external forces which depend on the flow velocity itself, such as Coriolis force \eqref{eq:coriolis}. In the latter case, algebraic equation \eqref{eq:trans_u1} has to be solved by iteration, cf. e. g. \cite{sawant_consistent_2021}.
Below, only cases with force independent of the flow velocity are considered for simplicity of presentation.}

For completeness, a detailed multi-scale analysis of the lattice Boltzmann equation  \eqref{eq:coll_stream} is presented in Appendix \ref{ap:CE_LBM}.
The analysis reveals that the shallow water equations \eqref{eq:shallow_water_h} and \eqref{eq:shallow_water} are recovered, and where the viscous stress tensor \eqref{eq:TNS} is endowed with the kinematic viscosity,
{\begin{equation}
\label{eq:nu_LB}
\nu={\left(\frac{1}{2\beta}-\frac{1}{2}\right)}\delta t\left(\frac{P_0}{h}\right),
\end{equation}
while the bulk viscosity remains $\eta$ thanks to the correction term \eqref{eq:correction}.
Comments are in order:
}
{
\begin{itemize}
\item As it is pertinent to a lattice Boltzmann formulation, the viscosity \eqref{eq:nu_LB} is decreasing with the increase of the relaxation parameter $\beta\to 1$, while the time step $\delta t$ remains fixed. With the above relation \eqref{eq:beta},
this so-called over-relaxation regime corresponds to smallness of the bare relaxation time $\tau$ relative to the time step, $\tau\ll \delta t$.
\item
{Finally, we note that, same as in the continuous time-space analysis, in the present lattice Boltzmann setting, the recovered hydrodynamic limit does not depend on the slow relaxation time $\lambda$, which to this end remained a free parameter. Based on the above comment, we set
\begin{equation}\label{eq:step}
    \lambda=\delta t.
\end{equation}
This is consistent with the requirement $\lambda\gg\tau$ \eqref{eq:dominance}, maintained in the over-relaxation regime $\beta\to 1$.}
\end{itemize}
}
This concludes the derivation of the proposed lattice Boltzmann framework for the simulation of shallow water equations. Below, we shall summarize the implementation. {In addition and for the sake of comparison, a summary of a conventional lattice Boltzmann model for shallow water equation is provided in Appendix \ref{app:on_diff}.}

\subsection{Summary of the lattice Boltzmann model}
{Populations $f_i$ evolve following equation Eq.~\eqref{eq:coll_stream}, where we drop the over-bar for readability and set $\lambda=\delta t$ in accord with \eqref{eq:step},}
\begin{equation}\label{eq:coll_stream_summ}
{f_i(\bm{x}+\bm{c}_i\delta t, t+\delta t) = f_i(\bm{x}, t) + 2\beta\left(f_i^{\rm eq}(h,\bm{u}) - f_i(\bm{x},t)\right) + 
\left(1-\beta\right) \left(f_i^{*}(h,\bm{u};\delta t) - f_i^{\rm eq}(h,\bm{u})\right).}
\end{equation} 
Evaluation of the equilibrium $f_i^{\rm eq}$ and of the shifted-equilibrium $f_i^*$ in Eq.\ \eqref{eq:coll_stream_summ} proceeds as follows:
\begin{itemize}
\item Functional dependence of the equilibrium populations $f_i^{\rm eq}$ on the water column height $h$ and on the flow velocity $\bm{u}$ is defined as in Eq.\ \eqref{eq:LBMeq} 
while that of the shifted-equilibrium populations $f_i^*$ is defined through Eqs.\ \eqref{eq:LBMstar}, \eqref{eq:u_shift} and \eqref{eq:correction}. 
\item The model comes with a degree of freedom in the form of the reference pressure $P_0$ which is chosen as in \eqref{eq:A} or \eqref{eq:B}. 
\item
In accord with Eqs.\ \eqref{eq:trans_h} and \eqref{eq:u_LBM}, the conserved moments are computed as,
\begin{align}
&    h(\bm{x},t) = \sum_{i=1}^Q f_i(\bm{x},t),\label{eq:mom_h_summ}\\
\label{eq:mom_u_summ}
&	   h \bm{u}(\bm{x},t) = \sum_{i=1}^Q \bm{c}_i f_i (\bm{x},t)+ \frac{\delta t}{2} \bm{F}(\bm{x},t).
\end{align}
\item The force  $\bm{F}$ in \eqref{eq:mom_u_summ} has the form \eqref{eq:u_shift} and involves a contribution with the space derivative of $P-P_0$.
The latter is evaluated with a finite difference method,
\begin{equation}
    \partial_{\alpha}(P-P_0){(\bm{x},t)} = \frac{1}{\delta t}\sum_{i=1}^Q \frac{w_i}{\varsigma^2}{c}_{i\alpha} \left(P(\bm{x}+\bm{c}_i\delta t{,t}) - P_0(\bm{x}+\bm{c}_i\delta t{,t})\right),\label{eq:discrete_F}
\end{equation}
{where $w_i$ are the weights of the discrete velocities of the D$2$Q$9$ lattice} \eqref{eq:D2Q9c},
\begin{equation}\label{eq:weights_2d}
    w_i = w_{c_{ix}} w_{c_{iy}},\ w_0=\frac{2}{3},\ w_1=w_{-1}=\frac{1}{6}.
\end{equation}
Note that the force contribution \eqref{eq:discrete_F} vanishes if the reference pressure is chosen as in Case B \eqref{eq:B}, $P_0=P$.

Once the water column height field $h$ is updated using Eq. \eqref{eq:mom_h_summ}, the derivative in  Eq. \eqref{eq:discrete_F} can be evaluated and, consequently, the velocity field $\bm{u}$ is updated {according to Eq.\ \eqref{eq:mom_u_summ}}. The updated velocity field and its space derivatives, featured in the correction term \eqref{eq:correction},
\begin{equation}
    \partial_\alpha u_\alpha {(\bm{x},t)} = \frac{1}{\delta t}\sum_{i=1}^Q \frac{w_i}{\varsigma^2} c_{i\alpha} u_\alpha(\bm{x}+\bm{c}_i\delta t{,t}),\label{eq:discrete_u}
\end{equation}
are then plugged into $f_i^{\rm eq}$ and $f_i^{*}$ and which concludes the evaluation of the latter.
\item 
The relaxation parameter $\beta\in[0,1]$ is tied to the kinematic viscosity $\nu$ \eqref{eq:nu_LB}.
\item The bulk viscosity $\eta$ is directly entering $f_i^*$ through Eq. \eqref{eq:correction}, where the relaxation time $\tau$ is set by inverting the relation \eqref{eq:beta},
\begin{equation}
    \label{eq:beta_to_tau}
    \tau=\left(\frac{1}{2\beta}-\frac{1}{2}\right)\delta t.
\end{equation}
\item Finally, the on-lattice propagation in the left-hand side of Eq.\ \eqref{eq:coll_stream_summ} is set by a requirement for the grid spacing, $\delta x=\sqrt{3}\varsigma \delta t$, where the lattice speed of sound $\varsigma$ is given by \eqref{eq:cs}.
\end{itemize}
The overall structure of the algorithm is shown in Fig. \ref{Fig:FlowChart}.
  \begin{figure}
    \hspace{0.5 cm}
  \begin{tikzpicture}[auto,
    block_m/.style ={rectangle, draw=black, thick, fill=white,
      text width=14em, text centered,
      minimum height=3em},
    block_s/.style ={rectangle, draw=black, thick, fill=white,
      text width=8em, text centered,
      minimum height=3em}]
    \node[block_m] at (0,1) (ini) {Initialize: $f_i=f_i^{\rm eq}$};
    
    \node[block_m] at (0,-0.5) (moments) {Compute $h$, $P_0(h)$, $\bm{\nabla}P_0$};
    \node[block_s] at (0,-3.5) (correction){Compute $\Phi_{\alpha\alpha}$: \eqref{eq:correction}};
    \node[block_s] at (3.2,-6.5)  (omega) {Compute $\beta$: Eqs.~\eqref{eq:beta} and \eqref{eq:visc_gen}.};
    \node[block_s] at (3.2,-2) (force) {Compute $F_{\alpha}$: \eqref{eq:u_shift}};
    \node[block_s] at (3.2,-3.5) (mom_u) {Compute $\bm{u}$: \eqref{eq:u_LBM}};
    \node[block_m] at (1.6,-5) (feq) {Compute $f_i^{\rm eq}$: \eqref{eq:LBMeq}};
    \node[block_m] at (-3.5,-6.5) (fext) {Compute $f_i^*$: \eqref{eq:LBMstar}};
    \node[block_m] at (1.6,-8) (CollStream) {Collision and streaming: \eqref{eq:coll_stream}};
    \node[block_s] at (-5,-4) (next) {$t=t+\delta t$};

    \draw [->, to path={-| (\tikztotarget)}] (fext) edge (CollStream);
    \draw [->, to path={-| (\tikztotarget)}] (omega) edge (CollStream);
    \draw[->] (feq) -- (CollStream);

    \draw [->, to path={-| (\tikztotarget)}] (moments) edge (force);
    \draw[->] (force) -- (mom_u);
    \draw [->] (moments) -- (correction);
    \draw[->] (mom_u) -- (correction);
    \draw[->] (ini) -- (moments);

    \draw [->, to path={-| (\tikztotarget)}] (correction) edge (fext);
    \draw [->, to path={-| (\tikztotarget)}] (moments) edge (fext);
    \draw [->, to path={-| (\tikztotarget)}] (force) edge (fext);

    \draw [->] 
    (mom_u.east) -- +(0.5,0) |- +(0.5,-1.5) -- (feq.east);
    \draw [->] 
    (moments.east) -- +(3,0) |- +(3,-4.5) -- (feq.east);
    \draw [->] 
    (moments.east) -- +(3.5,0) |- +(3.5,-6) -- (omega.east);
    \draw [->] 
    (CollStream.west) -- +(-6,0) |- +(-6,4) -- (next.west);
    \draw [->] 
    (next.north) -- +(0,3.75) |- +(5,3.75) -- (moments.north);
  \end{tikzpicture}
       \caption{Overall structure of the proposed algorithm for the simulation of shallow water equations.}
      \label{Fig:FlowChart}
  \end{figure}
In the next section, we proceed with the numerical validation of the proposed lattice Boltzmann solver.

\section{Numerical application}
\subsection{Dispersion and dissipation properties}
\subsubsection{Measuring normal modes propagation speed}
The target hydrodynamic limit admits two normal eigen-modes, 
\begin{equation}
    \label{eq:sound}
    c^\pm = u \pm \sqrt{gh},
\end{equation} 
see spectral analysis in Appendix \ref{ap:SWE_spectral}. As a first step, we measure the propagation speed of eigenmodes from simulations for different values of $h$ via a weak pressure front. The setup consists of a 1-D domain in $h$ in the right half and $h+\delta h$ in the left half of the domain, with $\delta h\ll h$. Here we set {$h=1 [m]$} and $\delta h= 0.01 [m]$. Once the simulation is started, the pressure front moves at a speed that should correspond to the speed of sound.
\begin{figure}
	\centering
	\includegraphics[width=0.4\linewidth]{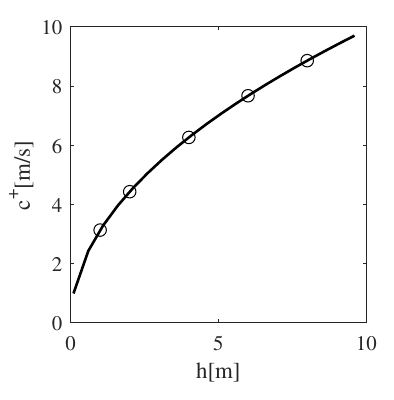}
	\caption{Sound speed as a function of column height. Line: Analytical, Eq. \eqref{eq:sound}; Symbol: Present lattice Boltzmann simulation with both the reference pressure $P_0=h\varsigma^2$ \eqref{eq:A} and $P_0=P$ \eqref{eq:B}. Results for both cases fall exactly on top of each other, only one set of data is shown.}
	\label{Figuresound}
\end{figure}
The results are shown in Fig. \ref{Figuresound} and point to excellent agreement with the analytical solution, confirming the consistency of the dispersion in the hydrodynamic limit.

\subsubsection{Galilean invariance of shear dissipation rate \label{sec:shear}}
Next, we model a plane shear wave to measure the kinematic viscosity. The wave corresponds to a small sinusoidal perturbation imposed on the initial velocity field. The initial conditions of the flows are,
\begin{equation}
    h(x,0) = 1{[m]},\ u_x(x,0) = u_{x0},\ u_y(x,0) = A\sin\left(2\pi x/L_x\right),
\end{equation}
where we set $A = 0.01~[m/s]$ and the domain size is $L_x = 10~[m]$. The periodic domain is discretized with grid points $200$. Simulations were performed for different values of $u_{x0}\in\left[-0.3, 0.3\right]~[m/s]$. The bulk viscosity is set to $\eta=0$ and $\beta=0.625$.
The apparent viscosity was measured by fitting the evolution of the maximum $y$-velocity in the domain to,
\begin{equation}
    u_x^{\rm max}\propto \exp{\left(-\frac{4\pi^2\nu}{L_x^2}t\right)}.
\end{equation}
\begin{figure}
	\centering
		\includegraphics[width=0.4\linewidth]{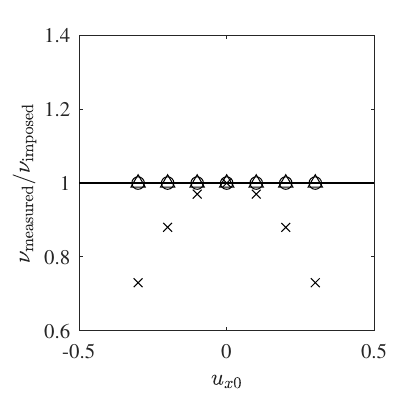}
	\caption{{Shear dissipation rate as obtained from simulation. Circular markers: Present model with the reference pressure $P_0=P$ \eqref{eq:B}; Triangular markers: Present model with  $P_0=h\varsigma^2$ \eqref{eq:A}; Cross markers: Standard LBGK \eqref{eq:coll_stream_LBGK} with Salmon's equilibrium \eqref{eq:salmon_equilibrium}.}}
	\label{Figureshear}
\end{figure}
{The measured viscous dissipation rates with the present lattice Boltzmann model and using both pressure splitting strategies \eqref{eq:A} and \eqref{eq:B} are presented in Fig. \ref{Figureshear}. For comparison, the result obtained with conventional LBGK \eqref{eq:coll_stream_LBGK} with Salmon's equilibrium \eqref{eq:salmon_equilibrium}, are also shown in Fig. \ref{Figureshear}.
While the dissipation rates of the proposed model with either pressure splitting show excellent agreement and demonstrate Galilean invariance of the shear dissipation rate in the hydrodynamic limit, i.e. for a well-resolved simulation, the conventional LBGK shows significant deviations at non-zero flow velocity.}

\subsubsection{Galilean invariance of normal dissipation rate \label{sec:normal}}
In order to measure effective dissipation rate of normal modes, a small perturbation was added to the $x$-component of the velocity field \citep{hosseini2020compressibility} generating perturbations that only affect the diagonal components of the viscous stress and placing the study in the linear regime. The flow was initialized as,
\begin{equation}\label{eq:initial_GI}
    h(x,0) = h_0,\ u_x(x,0) = u_{x0} + A\sin\left(2\pi x/L_x\right),\ u_y(x,0) = 0,
\end{equation}
where $A = 0.001~[m/s]$. {Various initial water column heights $h_0$ were probed.} We set $\eta=0.01~[m^2/s]$ and $\beta=0.625$. The decay of the maximum velocity relative to $u_{x0}$ over time was fitted to an exponential function depending on ${(\nu+\eta)}/{2}$ in 2-D,
\begin{equation}
    u_x^{\rm max} \propto \exp{\left(-\frac{2\pi^2(\nu+\eta)}{L_x^2}t\right)}.
\end{equation}

\begin{figure}
	\centering
		\includegraphics[width=0.4\linewidth]{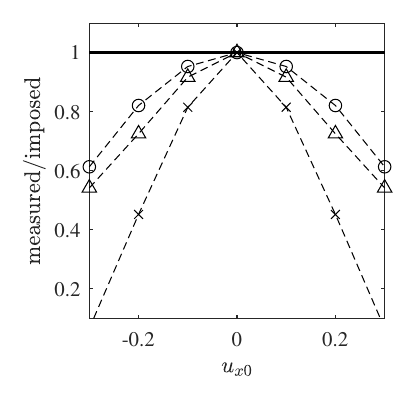}
	\caption{{Normal dissipation rates as obtained from simulations with the standard LBGK \eqref{eq:coll_stream_LBGK} using  Salmon equilibrium, Eq. \eqref{eq:salmon_equilibrium} for various water depth $h_0$ \eqref{eq:initial_GI}. Circular markers: $h_0=3~[m]$; Triangular markers: $h_0=2~[m]$; Cross markers: $h_0=1~[m]$.}}
	\label{normal_dissipation_rate_salmon}
\end{figure}

\begin{figure}
	\centering
		\includegraphics[width=0.8\linewidth]{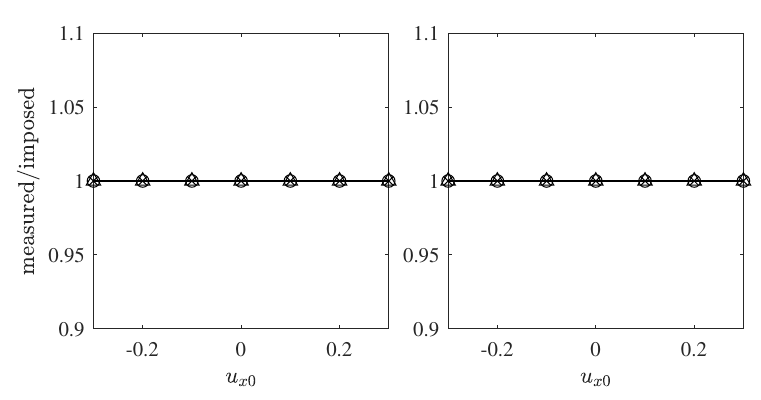}
	\caption{{Normal dissipation rates as obtained from simulations with the present model. Left: Reference pressure $P_0=P$ \eqref{eq:B}; Right: Reference pressure $P_0=h\varsigma^2$ \eqref{eq:A}. Circular markers: $h_0=3~[m]$; Triangular markers: $h_0=2~[m]$; Cross markers: $h_0=1~[m]$.}}
	\label{normal_dissipation_rate_new}
\end{figure}

We first investigate the effective dissipation rate for {standard LBGK \eqref{eq:coll_stream_LBGK} with }  Eq.~ \eqref{eq:salmon_equilibrium}. The results are shown in Fig. \ref{normal_dissipation_rate_salmon}.
The results point to violation of Galilean invariance in the normal dissipation rates: the effective dissipation rates of normal modes change with the flow speed. Furthermore, we observe that the difference in column height $h$ affects the Galilean invariance of the dissipation rates. This is related to the errors linear in the velocity in the diagonal components of the viscous stress tensor.
Similar simulations were carried out with the proposed scheme, with both cases of reference pressure, $P_0=P$ \eqref{eq:B} and $P_0=h\varsigma^2$ \eqref{eq:A}. The results are displayed in Fig.~\ref{normal_dissipation_rate_new}.
The proposed scheme allows to impose the intended normal dissipation rates and is Galilean invariant, i.e. dissipation rates are unaffected by the flow velocity.

\subsection{Flows with non-uniform bed topology}
\subsubsection{Static water column over a bump}
\begin{figure}
	\centering
		\includegraphics[width=0.4\linewidth]{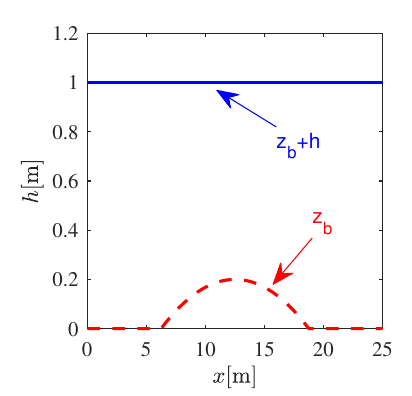}
	\caption{Initial conditions of static water over a bump test.}
	\label{bump_geo}
\end{figure}

\begin{figure}
	\centering
		\includegraphics[width=0.4\linewidth]{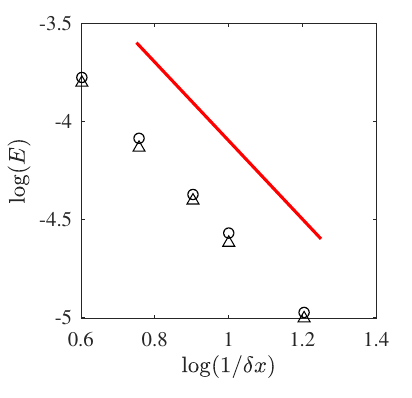}
	\caption{{Static water over a bump test: $L_2$-norm of the error at steady state. 
    Circle: Present model with $P_0=P$ \eqref{eq:B}; Triangle: Present model with $P_0=h\varsigma^2$ \eqref{eq:A}. 
    Solid red line shows the $-2$ slope.}}
	\label{bump_convergence}
\end{figure}

This configuration consists of a periodic 1-D domain of length $L_x=2~[m]$. The water bed is not uniform and has a profile with a bump,
\begin{equation}
    z_{\rm b} = \begin{cases}
0 & L_x/4> x \ {\rm or }\ 3L_x/4< x \\
0.2 - \frac{1.6}{L_x^2}{(x-L_x/2)}^2 & L_x/4\leq x \leq 3 L_x/4
\end{cases}.
\end{equation}
The initial velocity in the domain is set to zero while $h(x,t=0)=1-z_{\rm b}$. The initial conditions are shown in Fig. \ref{bump_geo}.
{At steady state, the analytical solution is the initial height profile. To assess the accuracy of the simulations, the system is left to evolve until steady state is reached. Simulations were performed with $\delta x\in\{0.25,0.172,0.125,0.1,0.06255\}~[m]$. The time step was set using $\delta x/\delta t=0.2~[m/s]$. Furthermore, $\eta=0.01\delta x^2/\delta t~[m/s]$ while $\beta=0.83$.
Steady results were compared to the analytical solution. The $L_2$-norm of the error is shown in Fig. \ref{bump_convergence}, pointing to a second-order accuracy of the present model for both pressure-splitting strategies, Case A \eqref{eq:A} and Case B \eqref{eq:B}.}

\subsubsection{Subcritical flow over a bump}
\begin{figure}
	\centering
		\includegraphics[width=0.4\linewidth]{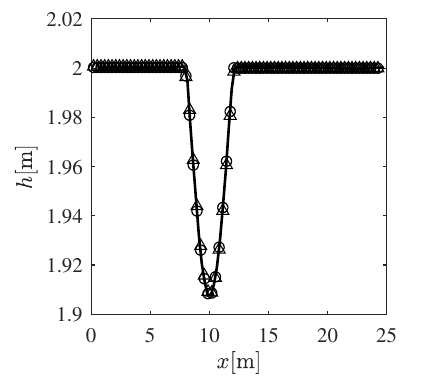}
	\caption{Subcritical flow over bump test: Circle and triangle markers are results obtained from the simulation with $P_0=P$ \eqref{eq:B} and $P_0=h\varsigma^2$ \eqref{eq:A}, respectively. Plain line is the analytical result from \cite{vazquez1999improved}. 
    }
	\label{bump_inlet_data}
\end{figure}

Next we turn our attention to an extension of the previous configuration proposed in \cite{vazquez1999improved}. 
The case consists of a 1-D domain of size $L_x=25~[m]$ subject to $h=2~[m]$ and $u=2.21~[m/s]$ on the left-hand side and outlet boundaries on the right-hand side. The inlet boundaries are imposed by setting the discrete distribution functions at the inlet node to equilibrium values while for the outlet, $x={\rm outlet}$, a simple zero-gradient approach is chosen where missing distribution function after streaming, $f_i$, are computed as,
\begin{equation}
    f_i(x, t) = f_i(x-\delta x, t).
\end{equation}
The bed shape is defined as,
\begin{equation}
    z_b = \begin{cases}
0 & 8> x \,{\rm or }\, 12< x \\
0.2 - 0.05{(x-10)}^2 & 8\leq x \leq 12
\end{cases}.
\end{equation}
while the domain is initialized with $h=2~[m]$ and $u=0$. The simulation was carried out with $\delta x=0.0625~[m]$ and $\delta t=0.00625~[s]$. Furthermore, $\eta=0.0125~[m^2/s]$ and $\beta=0.83$.
The steady-state results as obtained from the simulation are compared to the reference data from \cite{vazquez1999improved} in Fig. \ref{bump_inlet_data} and point to excellent agreement with both pressure splittings.
\subsection{1-D shock tube configurations}
\subsubsection{1-D Dam-break simulation}
\begin{figure}
	\centering
		\includegraphics[width=0.9\linewidth]{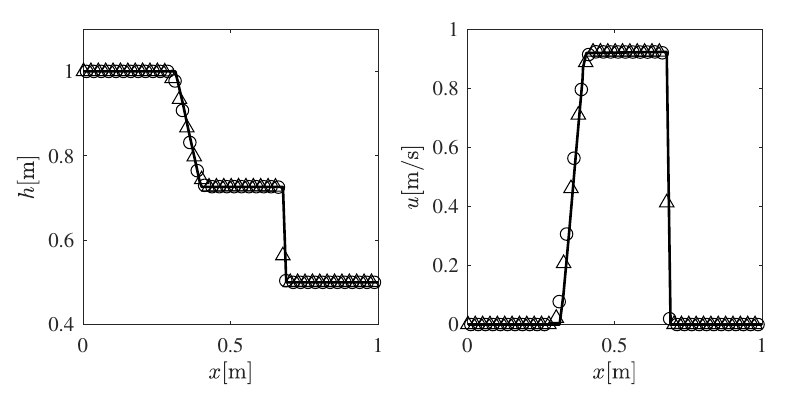}
	\caption{Dam-break case solution at $t=0.6~[s]$. Line represents analytical solution while circle and triangle markers are results obtained from simulation with $P_0=P$ \eqref{eq:B} and $P_0=h\varsigma^2$ \eqref{eq:A}, respectively.}
	\label{1ddam_plot}
\end{figure}

We move on to the validation of the proposed lattice Boltzmann model by considering a 1-D dam-break problem. The initial conditions for this test are as follows: the flow depths in the upstream and downstream half-domains are $1.0~[m]$ and $0.5~[m]$, respectively. The initial flow velocity is zero and the domain size is $1.0~[m]$. We ran simulations with both pressure splitting strategies with $\delta x=0.0025~[m]$. The time step is set using $\delta x/\delta t = 10~[m/s]$. Furthermore, $\eta=0.0125~[m^2/s]$ and $\beta=0.83$. The results obtained from the simulations are compared with the analytical solutions in Fig. \ref{1ddam_plot}.
The results of the simulation show excellent agreement with the analytical solution.
\subsubsection{Dam-break over step}
\begin{figure}
	\centering
		\includegraphics[width=0.9\linewidth]{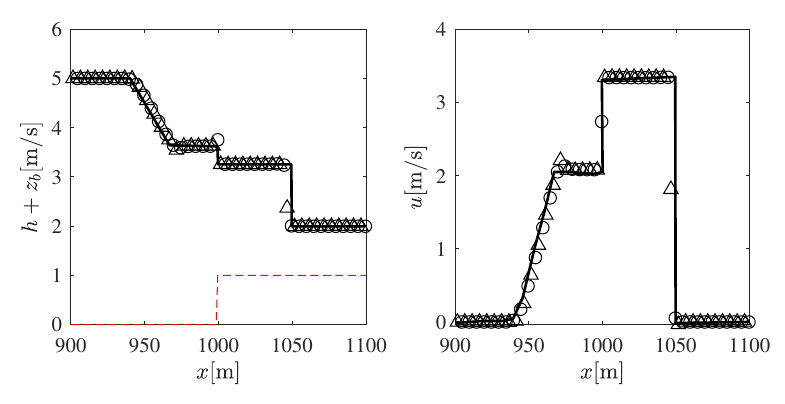}
	\caption{{Dam-break over a step at $t=0.6~[s]$. Left: Water column height; Right: Flow velocity. 
    Circle: Present model with $P_0=P$ \eqref{eq:B}; Triangle: Present model with $P_0=h\varsigma^2$ \eqref{eq:A}. Solid line: Reference analytical solution. For clarity, we show the data in $x\in[900,1100]~[m]$}. }
	\label{damstep}
\end{figure}

    The next case follows the setup proposed by \cite{rosatti2010riemann,begnudelli2011hyperconcentrated}. It consists of a 1-D domain of size $L_x=2000~[m]$ with the height of the water column in the left and right halves of the domain set to $5~[m]$ and $0.9966~[m]$, respectively. In addition, different from the previous case, the bed admits a step, with $z_{\rm b}=0$ on the left half of the bed and $z_{\rm b}=1~[m]$ on the right half of the domain. We run the simulations with $\delta x=1[m]$ and $\delta x/\delta t = 10~[m/s]$. Furthermore, $\eta=0.0125~[m^2/s]$ and $\beta=0.83$. The results are compared to the analytical solution in Fig. \ref{damstep}, and show excellent agreement.
\subsection{2-D cases}
\subsubsection{Circular dam break}
\begin{figure}
	\centering
		\includegraphics[width=0.8\linewidth]{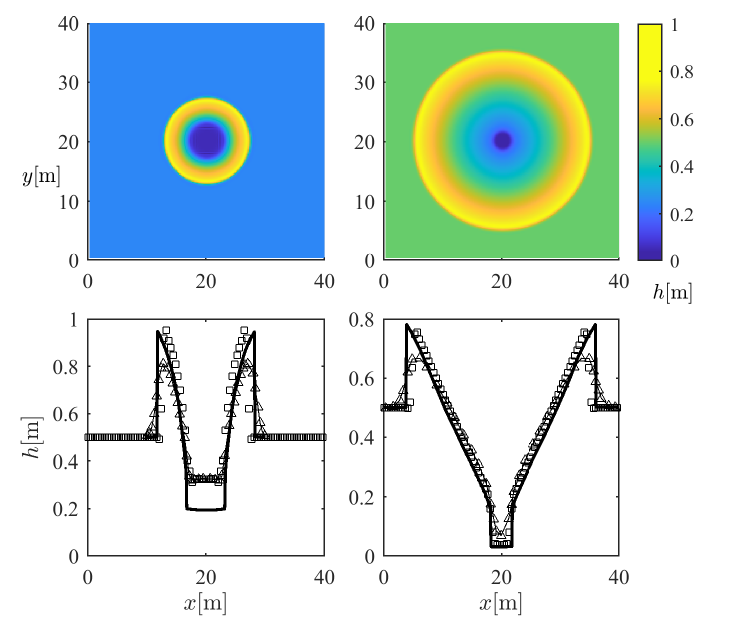}
	\caption{Results from simulation of circular dam break using the proposed solver with $P_0=P$ at times (left column) $t=1.2~[s]$ and (right column) $t=3.5~[s]$. Top row shows the water column hight fields. Bottom row shows height distribution along horizontal centerline. Plain lines: high resolution reference data from \cite{stecca2012finite}; Triangle markers: finite volume solver from \cite{stecca2012finite}; Square markers: present lattice Boltzmann simulation.}
	\label{circ_dam_pressure}
\end{figure}

\begin{figure}
	\centering
		\includegraphics[width=0.8\linewidth]{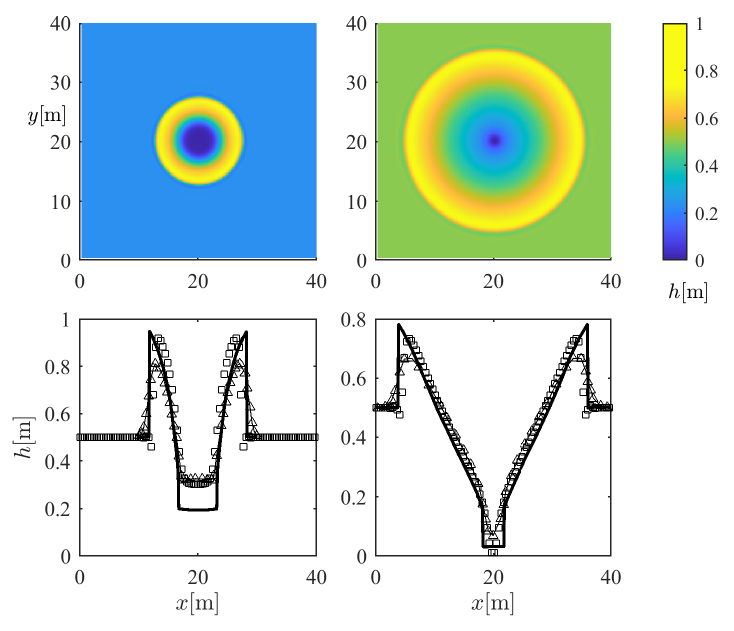}
	\caption{Results from simulation of circular dam break using the proposed solver with $P_0=h\varsigma^2$ at times (left column) $t=1.2~[s]$ and (right column) $t=3.5~[s]$. Top row shows the water column hight fields. Bottom row shows height distribution along horizontal centerline. Plain lines: high resolution reference data from \cite{stecca2012finite}; Triangle markers: finite volume solver from \cite{stecca2012finite}; Square markers: present lattice Boltzmann simulation.}
	\label{circ_dam_force}
\end{figure}

In this test, we examine the collapse of an idealized circular dam with a flat bottom. Simulating this scenario reveals a captivating time evolution of the resulting wave patterns. In addition, it provides an opportunity to assess the preservation of symmetry in the numerical solution. The problem was initially studied by \cite{alcrudo1993high}, and later by several others. The setup involves a cylindrical dam with a radius of $2.5~[m]$, positioned at the center of a square domain $40\times 40~[m^2]$. The initial water levels inside and outside the dam are $2.5~[m]$ and $0.5~[m]$, respectively. Simulations were carried out using both pressure splitting strategies over $t=3.5~[s]$. The simulation domain was discretized using $\delta x=0.4~[m]$ and $\delta x/\delta t = 10~[m/s]$. Furthermore, {the bulk viscosity} is set to a constant value of $\eta=0.05~[m^2/s]$ and $\beta=0.83$. The results are compared to reference data reported in \citep{stecca2012finite}, generated using an exact Godunov finite volume solver with both a resolution comparable to our simulations, i.e. $\delta x=0.4~[m]$ and a high-resolution simulation with $\delta x=0.04~[m]$. The results are shown in Fig. \ref{circ_dam_pressure} for the reference pressure \eqref{eq:B} and in Fig. \ref{circ_dam_force} for the reference pressure \eqref{eq:A}.

\begin{figure}
	\centering
		\includegraphics[width=0.8\linewidth]{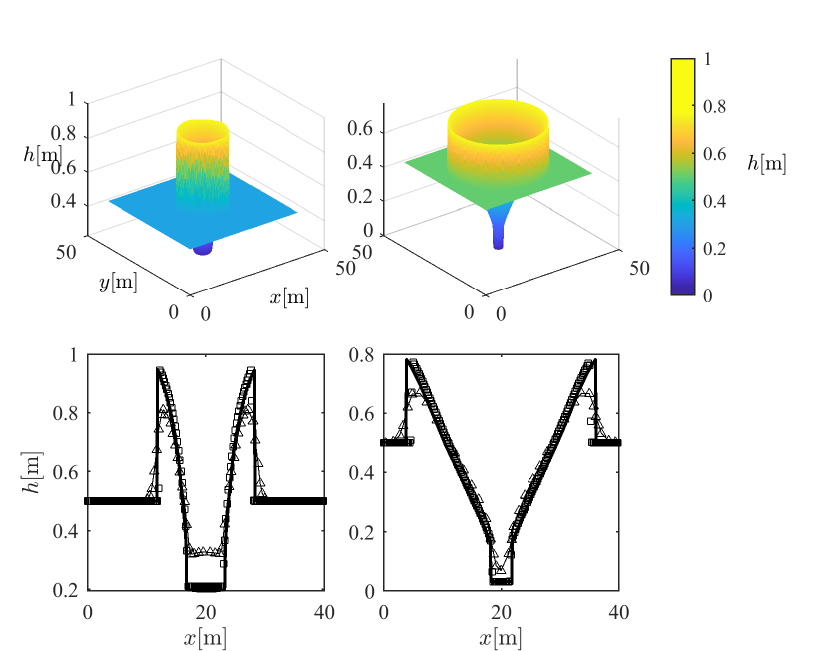}
	\caption{Results from high resolution simulation of circular dam break using the proposed solver with $P_0=h\varsigma^2$ at times (left column) $t=1.2~[s]$ and (right column) $t=3.5~[s]$. Top row shows the water column hight fields. Bottom row shows height distribution along horizontal centerline. Plain lines: high resolution reference data from \cite{stecca2012finite}; Triangle markers: finite volume solver from \cite{stecca2012finite}; Square markers: present lattice Boltzmann simulation.}
	\label{circ_dam_HR}
\end{figure}

We observe that both realizations of the pressure splitting are stable and maintain cylindrical symmetry, indicating a good isotropy behavior. Furthermore, the results are in very good agreement with reference solutions and show sharper peaks and less numerical dissipation than the Godunov solver. To show convergence, an additional simulation with $\delta x=0.04~[m]$ was also carried out. The results are shown in Fig. \ref{circ_dam_HR}.
The comparison points to excellent agreement with the reference solution.
\subsubsection{2-D partial dam break\label{section:2d_dam}}
\begin{figure}
	\centering
		\includegraphics[width=0.4\linewidth]{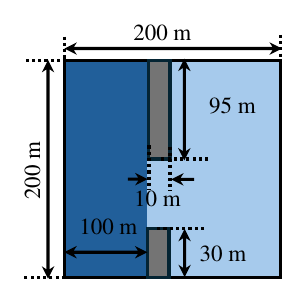}
	\caption{Geometry of partial dam break configuration.}
	\label{part_dam_geo}
\end{figure}

\begin{figure}
	\centering
		\includegraphics[width=0.8\linewidth]{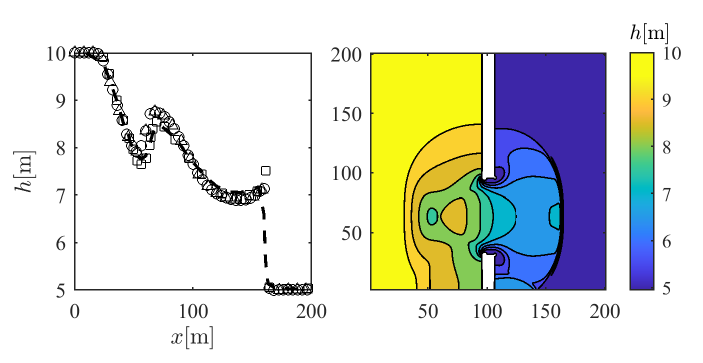}
	\caption{Left: Water height profile along the horizontal line passing through the center of the dam breach. Dashed line: finite-volume solver RiverFlow 2D from \cite{venturi2020new}; Circle and triangle markers: present simulation with $P_0=P$ and $P_0=h\varsigma^2$, respectively. Rectangle markers: a cumulant lattice Boltzmann simulation of \cite{venturi2020new}.
    Right: Iso-contours of water height. Lines are spaced with $\delta h=0.5~[m]$. Both plots from $t=7.2~[s]$.}
	\label{partial_dam_ref}
\end{figure}

\begin{figure}
	\centering
		\includegraphics[width=0.9\linewidth]{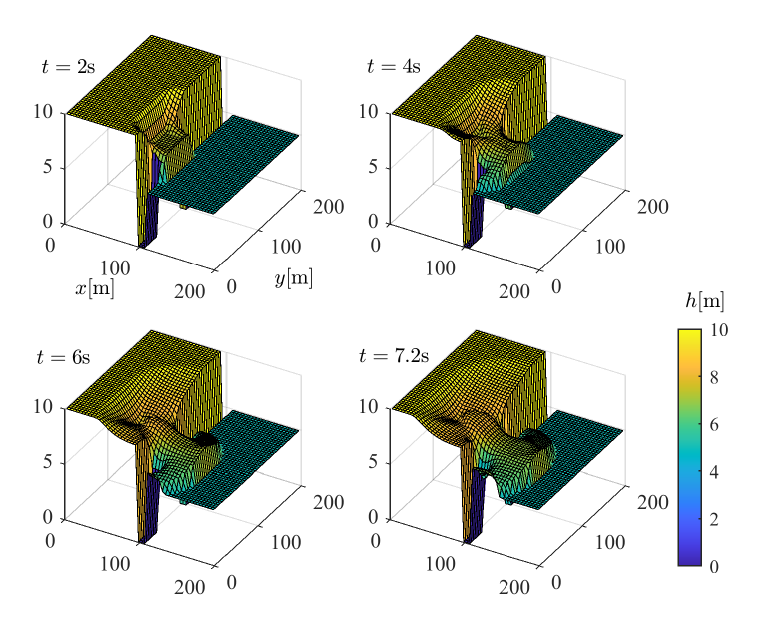}
	\caption{Snapshots of the water height field for the partial dam break configuration.}
	\label{partial_dam_snaps}
\end{figure}

{Finally, we consider the case of a partial dam breach due to a structural failure or instantaneous opening of sluice gates. This configuration was first studied by \cite{fennema1990explicit}. The case consists of a square domain of size $200\times200~[{m}^2]$, divided into left and right halves
by a wall of thickness $10~[m]$. The initial water levels in the left and right halves are $h_{\rm left}$ and $h_{\rm right}$, respectively.
At $t=0$, a partial dam failure is modeled by removing $75~[m]$ of the dividing wall.}
The geometry is shown in Fig. \ref{part_dam_geo}. 
Here, we run the configuration with $h_{\rm left}=10[m]$ and $h_{\rm right}=5~[m]$. The simulations are carried out over $8~[s]$, with $\delta x=0.5~[m]$ and $\delta t=0.05~[s]$. Furthermore, the bulk viscosity was set to $\eta=0.01~[m^2/s]$ and $\beta=0.83$. The results are shown in Fig. \ref{partial_dam_ref} and compared to the data reported in \cite{venturi2020new} and generated using a finite-volume solver for the shallow-water equation, RiverFlow 2D.
The comparison shows good agreement with the finite-volume solver. The evolution of the height of the water column over time is illustrated in Fig. \ref{partial_dam_snaps}.
{Furthermore, results obtained by \cite{venturi2020new} using a cumulant lattice Boltzmann model are also shown in Fig. \ref{partial_dam_ref} for comparison.}
It is interesting to note in passing that the resolution employed in the present simulation, $\delta x=0.5~[m]$, is $1/5$ of that used in \citep{venturi2020new}, $\delta x=0.1~[m]$, while the time step size used here, that is, $\delta t=0.05~[s]$ is approximately six times that in \citep{venturi2020new}, i.e. $\delta t=0.0082~[s]$. 

\section{\label{sec:level4}Conclusion} 
A novel family of discrete velocity Boltzmann equations models was introduced that focuses on the hydrodynamics of viscous shallow-water equations. The model relies on minimalist D$2$Q$9$ lattice with proper correction restoring independence of normal modes dissipation rates and full Galilean invariance of dissipation rates at the Navier-Stokes level. Furthermore, a novel interpretation of force contributions as a relaxation term leading to a general and consistent way to introduce them into the kinetic equations was proposed. In addition, the model allowed for splitting the pressure contribution at will between the collision operator and a force-like contribution, with the hydrodynamic level dynamics remaining invariant. Subsequently, a consistent on-lattice Lagrangian discretization in space and time was introduced. The consistent discretization led to a second-order scheme with no errors stemming from the force term present at the Euler and Navier-Stokes levels. Numerical applications showed that the proposed family of solvers properly recovers dispersion and dissipation at the hydrodynamic level and is stable and accurate when used for complex configurations.

\section*{Acknowledgement}
This work was supported by European Research Council (ERC) Advanced Grant No. 834763-PonD and by the Swiss National Science Foundation (SNSF) Grants 200021-228065 and 200021-236715. Computational resources at the Swiss National Super Computing Center (CSCS) were provided under Grants No. s1286 and sm101. 

\section*{Declaration of interests}
The authors report that they do not have a conflict of interest.

\section*{Data Availability Statement}
The data that support the findings of this study are available from the corresponding author upon request.

\appendix
\section{Chapman--Enskog analysis of the kinetic model\label{ap:CE_DVBE}}
We start with the discrete velocity system of equations shown in Eq.~\eqref{eq:final_kinetic_model},
\begin{equation}\label{eq:final_kinetic_model_app}
	\partial_t f_i + \bm{c}_i\cdot\bm{\nabla} f_i = -\frac{1}{\tau}\left(f_i-f_i^{\rm eq}(h,\bm{u})\right) 	+ \frac{1}{\lambda}(f_i^*(h,\bm{u};\lambda)-f_i^{\rm eq}(h,\bm{u})).
\end{equation}
We introduce the following parameters: characteristic flow velocity $\mathcal{U}$, characteristic flow scale $\mathcal{L}$, characteristic flow time $\mathcal{T}=\mathcal{L}/\mathcal{U}$ and characteristic water column height $\bar{h}$.
With the above, reduced variables are as follows: time $t=\mathcal{T}t'$, space $\bm{x}=\mathcal{L}\bm{x}'$, flow velocity $\bm{u}=\mathcal{U}\bm{u}'$, discrete particle velocity {$\bm{c}_i=c_s\bm{c}_i'$} where $c_s=\sqrt{g\bar{h}}$,
water column height $h$=$\bar{h} h'$, {distribution function $f_i=\bar{h}f_i'$.}
In addition we also introduce the following non-dimensional groups: Knudsen number ${\rm Kn}={\tau c_s}/{\mathcal{L}}$ and Mach number ${\rm Ma}={\mathcal{U}}/{c_s}$. This leads to,
\begin{equation}\label{eq:final_kinetic_model_nondim}
	[{\rm Kn}][{\rm Ma}]\partial_{t'} f'_i + [{\rm Kn}]\bm{c}'_i\cdot\bm{\nabla}' f'_i = -\left(f'_i-{f_i^{\rm eq}}'\right) + \frac{\tau}{\lambda}({f_i^*}'-{f_i^{\rm eq}}').
\end{equation}
We then assume the following scalings: Acoustic scaling, ${\rm Ma}\sim 1$ and Hydrodynamic scaling ${\rm Kn} \sim\epsilon$. We also assume $\tau/\lambda\le 1$. This leads to the following non-dimensional system of equations,
\begin{equation}\label{eq:final_kinetic_model_nondim_1}
	{\epsilon}\partial_{t'} f'_i + {\epsilon}\bm{c}'_i\cdot\bm{\nabla}' f'_i = -\left(f'_i-{f_i^{\rm eq}}'\right) + \frac{\tau}{\lambda} ({f_i^*}'-{f_i^{\rm eq}}').
\end{equation}
With the non-dimensional system of equations in hand, to analyze the balance equations recovered in the hydrodynamic limit we expand variables as power series of the smallness parameter $\epsilon$ {(primes will be dropped to ease notation)},
	\begin{eqnarray}
        \partial_t &=& \partial_t^{(1)} + \epsilon \partial_t^{(2)} + O(\epsilon^2),\\
	    f_i &=& f_i^{(0)} + \epsilon f_i^{(1)} + \epsilon^2 f_i^{(2)} + O(\epsilon^3),\\
	    {f^*_i} &=& {f^*_i}^{(0)} + \epsilon {f^*_i}^{(1)} + \epsilon^2 {f^*_i}^{(2)} + O(\epsilon^3).
	\end{eqnarray}
Introducing these expansions and separating terms of different orders, $\epsilon^n$ with $n\in\{0,1,2\}$:
\begin{subequations}
	\begin{align}
    \epsilon^0 &: {f^*_i}^{(0)} = f_i^{\rm eq},\\
    \epsilon &: f_i^{(0)} = f_i^{\rm eq},\\
	\epsilon^2 &: \mathcal{D}_{t}^{(1)} f_i^{(0)} = -\frac{1}{\tau} f_i^{(1)} + \frac{1}{\lambda}{f^*}_i^{(1)},\\
	\epsilon^2 &: \partial_t^{(2)}f_i^{(0)} + \mathcal{D}_{t}^{(1)} f_i^{(1)} = -\frac{1}{\tau} f_i^{(2)} + \frac{1}{\lambda}{f^*}_i^{(2)},
	\end{align}
    \label{Eq:CE_Eq_orders}
\end{subequations}
where $\mathcal{D}_{t}^{(1)} = \partial_t^{(1)} + \bm{c}_i\cdot\bm{\nabla}$. The following solvability conditions apply,
    \begin{eqnarray}
		\sum_{i=1}^Q {f}_i^{(k)} &=& 0, \forall k>0,\label{Eq:CE_solvability_DVM_1}\\
        \sum_{i=1}^Q \bm{c}_i {f}_i^{(k)} &=& 0, \forall k>0.\label{Eq:CE_solvability_DVM_2}
    \end{eqnarray}
The moments of the non-local contributions are:
    \begin{eqnarray}
		\sum_{i=1}^Q {f^*}_i^{(k)} &=& 0, \forall k>0,\\
		\sum_{i=1}^Q \bm{c}_{i} {f^*_i}^{(1)} &=& \lambda\bm{F},\\
		\sum_{i=1}^Q \bm{c}_{i}\otimes\bm{c}_{i} {f^*_i}^{(1)} &=& \lambda(\bm{u}\otimes\bm{F} + {\bm{F}\otimes\bm{u}}) + \lambda\Phi\bm{I},\\
        \sum_{i=1}^Q \bm{c}_{i}\otimes\bm{c}_{i} {f^*_i}^{(2)} &=& \frac{\lambda^2}{h}\bm{F}\otimes\bm{F}.
	    \label{Eq:CE_solvability}
    \end{eqnarray}
Taking the moments of the Chapman-Enskog-expanded equation at order $\varepsilon$:
	\begin{eqnarray}
	    \partial_t^{(1)}h + \bm{\nabla}\cdot h \bm{u} &=& 0,\label{eq:approach2_continuity1}\\
	    \partial_t^{(1)}h \bm{u} + \bm{\nabla}\cdot h \bm{u}\otimes\bm{u} + \bm{\nabla}P_0 &=& \bm{F},\label{eq:approach2_NS1}
	\end{eqnarray}
where $\bm{F}=\bm{\nabla}(P_0-P) + \bm{F}_{\rm ext}$. We note that the second relaxation time $\lambda$ does not appear at the Euler level and cancels out.\\
At order $\varepsilon^2$ the continuity equation is:
	\begin{equation}
	    \partial_t^{(2)} h = 0.\label{eq:approach2_continuity2}
	\end{equation}
For the momentum equations we have:
	\begin{equation}\label{eq:eps2_mom1_1}
	    \partial_t^{(2)} h \bm{u}
	    - \bm{\nabla}\cdot\tau\left[\partial_t^{(1)}\Pi_{2}^{(0)}+\bm{\nabla}\cdot\Pi_{3}^{(0)} - \bm{u}\otimes\bm{F} - \bm{F}\otimes\bm{u} - \Phi\right] = 0,
	\end{equation}
where $\Pi_{2}^{(0)}$ and $\Pi_{3}^{(0)}$ are the second- and third-order moments of $f_i^{(0)}$ defined as:
    \begin{eqnarray}
        \Pi_{2}^{(0)} &=& h \bm{u}\otimes\bm{u} + P_0\bm{I},\\
        \Pi_{3}^{(0)} &=& \Pi_{3}^{\rm MB} - \bm{u}\otimes \left(h\bm{u} \otimes\bm{u} + 3(P_0 - h \varsigma^2)\bm{I}\right)\circ\bm{J}
    \end{eqnarray}
where $\Pi_{\alpha\beta\gamma}^{\rm MB}=h u_\alpha u_\beta u_\gamma + P_0{\rm perm}(u_\alpha \delta_{\beta\gamma})$ is the third-order moment of the Maxwell-Boltzmann distribution, and for the sake of simplicity we have introduced the diagonal rank three tensor $\bm{J}$, with $J_{\alpha\beta\gamma}=\delta_{\alpha\beta}\delta_{\alpha\gamma}\delta_{\beta\gamma}$ and $\circ$ is the Hadamard product.
The contributions from the time derivative of the equilibrium second order moments tensor can be expanded as:
    \begin{align}
	\partial_t^{(1)} \Pi_{2}^{(0)}=&
	 \partial_t^{(1)}h \bm{u}\otimes\bm{u} + \partial_t^{(1)} P_0\bm{I}\nonumber\\
	 = & \bm{u}\otimes\partial_t^{(1)}h \bm{u} + {(\partial_t^{(1)}h \bm{u})\otimes\bm{u}} - \bm{u}\otimes\bm{u} \partial_t^{(1)} h + \partial_t^{(1)} P_0 \bm{I}\nonumber\\
	=& -\bm{\nabla}\cdot h  \bm{u}\otimes\bm{u}\otimes\bm{u} 
	    -\left[\bm{u}\otimes\left(\bm{\nabla} P_0 - \bm{F}\right) + {\left(\bm{\nabla} P_0 - \bm{F}\right)\otimes\bm{u}}
	     \right] + \partial_t^{(1)}P_0 \bm{I}
    \end{align}
and:
	\begin{multline}
	    \bm{\nabla}\cdot\Pi_{3}^{(0)} = \bm{\nabla}\cdot h \bm{u}\otimes\bm{u}\otimes\bm{u} 
	    + \left(\bm{\nabla} P_0 \bm{u} + {\bm{\nabla} P_0 \bm{u}}^{\dagger}\right) + (\bm{\nabla}\cdot P_0\bm{u})\bm{I}\\
	    - \bm{\nabla}\cdot\left[\bm{u}\otimes \left(h\bm{u} \otimes\bm{u} + 3(P_0 - h \varsigma^2)\right)\circ\bm{J}\right],
	\end{multline}
resulting in:
	\begin{multline}\label{eq:2nd_order_1}
	    \partial_t^{(1)} \Pi_{2}^{(0)} + \bm{\nabla}\cdot\Pi_{3}^{(0)} = P_0\left(\bm{\nabla}\bm{u} + {\bm{\nabla}\bm{u}}^{\dagger} \right) + \left(\bm{u}\otimes\bm{F} + {\bm{u}\otimes\bm{F}}^{\dagger}\right)\\
	    + \left(\bm{\nabla}\cdot P_0 \bm{u} + \partial_t^{(1)}P_0\right)\bm{I} - \bm{\nabla}\cdot\left[\bm{u}\otimes \left(h\bm{u} \otimes\bm{u} + 3(P_0 - h \varsigma^2)\right)\circ\bm{J}\right].
	\end{multline}
Plugging this last equation back into Eq.\ \eqref{eq:eps2_mom1_1}:
	\begin{multline}\label{eq:eps2_mom1_2}
	    \partial_t^{(2)}h \bm{u} - \bm{\nabla}\cdot\tau P_0\left(\bm{\nabla}\bm{u} + {\bm{\nabla}\bm{u}}^{\dagger}\right)
	    - \bm{\nabla}\tau \left(\partial_t^{(1)}P_0+\bm{\nabla}\cdot P_0 \bm{u}\right) \\- \bm{\nabla}\cdot\tau\left[\bm{\nabla}\cdot\left(\bm{u}\otimes \left(h\bm{u} \otimes\bm{u} + 3(P_0 - h \varsigma^2)\right)\circ\bm{J}\right) - \Phi\right] = 0.
	\end{multline}
Next we can expand the material derivative of $P_0$ as,
    \begin{equation}
        \partial_t P_0 + \bm{\nabla}\cdot P_0 \bm{u} =
        \partial_h P_0\left(\partial_t h + \bm{\nabla}\cdot h\bm{u}\right) + \left(P_0 - h\partial_h P_0\right)\bm{\nabla}\cdot\bm{u},
    \end{equation}
where $P_0=P_0(h)$. Using the balance of $h$, this equation simplifies to,
    \begin{equation} \label{eq:pressure_balance}
    \partial_t P_0 + \bm{\nabla}\cdot P_0 \bm{u} =\\ 
        \left(P_0 - h\partial_h P_0\right)\bm{\nabla}\cdot\bm{u}.
    \end{equation}
We now plug in $\Phi$ defined as,
	\begin{equation}\label{eq:corr_app}
	    \Phi_{\alpha\alpha} = \partial_{\alpha}\left(h u_{\alpha} \left(u_{\alpha}^2 + \frac{3P_{0}}{h}-3\varsigma^2\right)\right) + \left[\frac{\eta}{\tau} - P_0\left(\frac{D+2}{D}-\frac{\partial \ln P_0}{\partial \ln h}\right)\right] \bm{\nabla}\cdot\bm{u},
	\end{equation}
and Eq.~\eqref{eq:pressure_balance} into Eq.~\eqref{eq:eps2_mom1_2},
	\begin{equation}\label{eq:eps2_mom1_3}
	    \partial_t^{(2)}h \bm{u} - \bm{\nabla}\cdot\left[\tau P_0\left(\bm{\nabla}\bm{u} + {\bm{\nabla}\bm{u}}^{\dagger} - \frac{2}{D}\bm{\nabla}\cdot\bm{u}\bm{I}\right)
	    + h\eta\bm{\nabla}\cdot\bm{u}\bm{I}\right] = 0.
	\end{equation}
Tying the relaxation time $\tau$ to kinematic viscosity as,
\begin{equation}
    h\nu = \tau P_0,
\end{equation}
the following momentum balance equation at order $\epsilon^2$ is recovered:
\begin{equation}
    \partial_t^{(2)}h\bm{u} + \bm{\nabla}\cdot\bm{T}_{\rm NS} = 0.
\end{equation}
where $\bm{T}_{\rm NS}$ is given in Eq.~\eqref{eq:TNS}. Adding up orders $\epsilon$ and $\epsilon^2$ the following mass and momentum balance equations are recovered:
\begin{equation}
    \partial_t h + \bm{\nabla}\cdot h\bm{u} + \mathcal{O}(\epsilon^3)= 0,
\end{equation}
and,
\begin{equation}
    \partial_t h\bm{u} + \bm{\nabla}\cdot h\bm{u}\otimes\bm{u} + \bm{\nabla} P + \bm{\nabla}\cdot\bm{T}_{\rm NS} + \mathcal{O}(\epsilon^3) = \bm{F}_{\rm ext}.
\end{equation}

In order to illustrate the above asymptotic analysis, consider, for instance, the simulation of section \ref{section:2d_dam} with the parameters: $\beta=0.83$, leading to $\tau=0.005~[s]$, $\lambda = \delta t = 0.05~[s]$, $\mathcal{L}= \delta x = 0.5~[m]$, and $c_s= \sqrt{g\bar{h}} = 8.6~[m/s]$ where $\bar{h}=7.5~[m]$ and $g=9.81~[m/s^2]$. With this, 
we recover $\epsilon=0.086$, which is consistent with $\epsilon\ll 1$ and $\tau <\lambda$. 

\section{Chapman-Enskog analysis of the lattice Boltzmann model\label{ap:CE_LBM}}
For the lattice Boltzmann model, the first step in the multi-scale analysis is a Taylor expansion around $(\bm{x},t)$, leading to the following space and time-evolution equations,
\begin{equation}
    \delta t\mathcal{D}_t f_i + \frac{\delta t^2}{2}{\mathcal{D}_t}^2 {f_i}+ \mathcal{O}\left(\delta t^3\right) = 2\beta\left(f_i^{\rm eq} - f_i\right) + \frac{\delta t}{\lambda}\left(1-\beta\right)\left(f_i^{*} - f_i^{\rm eq}\right).
\end{equation}
Introducing the flow characteristic size and time, $\mathcal{L}$ and $\mathcal{T}$ the equation is made non-dimensional as,
\begin{equation}
    \frac{\delta x}{\mathcal{L}}\mathcal{D}'_t f_i + \frac{\delta x^2}{2\mathcal{L}^2}{\mathcal{D}'_t}^2 {f_i} = 2\beta\left(f_i^{\rm eq} - f_i\right) + \frac{\delta t}{\lambda}\left(1-\beta\right)\left(f_i^{*} - f_i^{\rm eq}\right),
\end{equation}
where,
\begin{equation}
    \mathcal{D}'_t = \frac{\mathcal{L}/\mathcal{T}}{\delta x/\delta t}\left(\partial'_t + \bm{c}'_i\cdot\bm{\nabla}'\right).
\end{equation}
Assuming acoustic scaling and hydrodynamic scaling, {$\varepsilon\sim\delta x/\mathcal{L} \sim\delta t/\mathcal{T}$} and dropping the primes,
\begin{equation}
    \varepsilon\mathcal{D}_t f_i + \frac{\varepsilon^2}{2}{\mathcal{D}_t}^2{f_i} = 2\beta\left(f_i^{\rm eq} - f_i\right) + \frac{\delta t}{\lambda}\left(1-\beta\right)\left(f_i^{*} - f_i^{\rm eq}\right),
\end{equation}
Introducing multi-scale expansions and separating in orders of the smallness parameter,
	\begin{subequations}
	\begin{align}
    \varepsilon^0 &: f_i^{(0)} = f_i^{\rm eq},\\
	\varepsilon &: \mathcal{D}_{t}^{(1)} f_i^{(0)} = -2\beta f_i^{(1)} + \frac{\delta t}{\lambda}\left(1-\beta\right){f^*}_i^{(1)},\\
	\varepsilon^2 &: \partial_t^{(2)}f_i^{(0)} + \mathcal{D}_{t}^{(1)} f_i^{(1)} + \frac{1}{2}{\mathcal{D}_{t}^{(1)}}^2 f_i^{(0)}= -2\beta f_i^{(2)} + \frac{\delta t}{\lambda}\left(1-\beta\right){f^*}_i^{(2)}.
	\end{align}
    \label{Eq:CE_Eq_orders_LB}
    \end{subequations}
One other point to note here is that the change of variables leads to the following solvability conditions,
    \begin{eqnarray}
		\sum_{i=1}^Q {f}_i^{(k)} = 0,\ \forall k>0,\\
		\sum_{i=1}^Q \bm{c}_{i} {f^{(1)}_i}  + \frac{\delta t}{2\lambda}\sum_{i=1}^Q \bm{c}_{i} f^{*(1)}_i =0,\label{Eq:CE_solvability_LB_2}\\
        \sum_{i=1}^Q \bm{c}_i {f}_i^{(k)} = 0,\ \forall k>1.
	    \label{Eq:CE_solvability_LB_3}
    \end{eqnarray}
The solvability condition \eqref{Eq:CE_solvability_LB_2} is obtained starting with Eq.~\eqref{Eq:CE_solvability_DVM_2} and plugging in Eq.~\eqref{eq:mom_u_summ}.
Now taking the zeroth- and first-order moments at order $\varepsilon$,
	\begin{eqnarray}
	    \partial_t^{(1)}h + \bm{\nabla}\cdot h \bm{u} &=& 0,\label{eq:approach2_continuity1_LB}\\
	    \partial_t^{(1)}h \bm{u} + \bm{\nabla}\cdot h \bm{u}\otimes\bm{u} + \bm{\nabla}P_0 &=& {-}2\beta \underbrace{\sum_{i=1}^Q \bm{c}_i\left(f_i^{(1)} + \frac{\delta t}{2\lambda}{f_i^*}^{(1)}\right)}_{=0} + \underbrace{\frac{\delta t}{\lambda}\sum_{i=1}^Q \bm{c}_i {f_i^*}^{(1)}}_{\bm{F}},\label{eq:approach2_NS1_LBM}
	\end{eqnarray}
At order $\varepsilon^2$, the continuity equation is obtained as,
\begin{equation}
\partial_t^{(2)} h = 0.
\end{equation}
while for the momentum equation,
\begin{equation}
    \partial_t^{(2)}h\bm{u} + \bm{\nabla}\cdot\left[\left(\frac{1}{2}-\frac{1}{2\beta}\right)\left(\partial_t^{(1)}\Pi^{(0)}_2 + \bm{\nabla}\cdot\Pi^{(0)}_3 - \frac{\delta t}{\lambda}\sum_{i=1}^Q \bm{c}_i\otimes\bm{c}_i {f_i^{*}}^{(1)}\right) \right] = 0,
\end{equation}
where using ~\eqref{eq:2nd_order_1},
\begin{equation}
    \left(\partial_t^{(1)}\Pi^{(0)}_2 + \bm{\nabla}\cdot\Pi^{(0)}_3 - \frac{\delta t}{\lambda}\sum_{i=1}^Q \bm{c}_i\otimes\bm{c}_i {f_i^{*}}^{(1)}\right) = P_0\left(\bm{\nabla}\bm{u} + \bm{\nabla}\bm{u}^\dagger - \frac{2}{D}\bm{\nabla}\cdot\bm{u}\bm{I}\right) + \frac{\eta}{1/2\beta-1/2}\bm{\nabla}\cdot\bm{u},
\end{equation}
which then leads to,
\begin{equation}
    \partial_t^{(2)}h\bm{u} + \bm{\nabla}\cdot\left[\left(\frac{1}{2}-\frac{1}{2\beta}\right)P_0\left(\bm{\nabla}\bm{u} + \bm{\nabla}\bm{u}^\dagger - \frac{2}{D}\bm{\nabla}\cdot\bm{u}\bm{I}\right) -h\eta \bm{\nabla}\cdot\bm{u}\bm{I}\right] = 0.
\end{equation}
Defining,
\begin{equation}
    \beta = \frac{\delta t}{\frac{2h\nu}{P_0} + \delta t},
\end{equation}
We recover,
\begin{equation}
    \partial_t^{(2)}h\bm{u} + \bm{\nabla}\cdot\bm{T}_{\rm NS} = 0,
\end{equation}
where $\bm{T}_{\rm NS}$ is defined in Eq.~\eqref{eq:TNS}.
Adding up the two orders in $\varepsilon$ we get the shallow water equations, i.e. Eqs.~\eqref{eq:shallow_water_h} and \eqref{eq:shallow_water}.\\
\section{On the difference with other models in the literature\label{app:on_diff}}
A review of the literature shows that equilibrium populations used for the LBM simulation of shallow water equations pertain to one of the following categories: (a) standard second-order polynomial equilibrium 
and (b) second-order polynomial supplemented by modifications to the fourth-order contracted ghost moment \citep{salmon1999lattice}. The latter, referred to as Salmon's equilibrium in this article, is most widely used. Both cases (a) and (b) recover the same leading-order hydrodynamics. We will therefore only consider Salmon's equilibrium. Neglecting external body forces for the sake of simplicity, the standard lattice Bhatnagar--Gross--Krook (LBGK) model reads,
\begin{equation}\label{eq:coll_stream_LBGK}
f_i(\bm{x}+\bm{c}_i\delta t, t+\delta t)= f_i(\bm{x}, t) + 2\beta\left(f_i^{\rm eq}(\bm{x},t) - f_i(\bm{x},t)\right),
 \end{equation}
where Salmon's equilibrium is defined as \citep{salmon1999lattice},
\begin{equation}\label{eq:salmon_equilibrium}
    f_i^{\rm eq} = w_i h\left(1+\frac{\bm{c}_i\cdot\bm{u}}{\varsigma^2}+\frac{1}{2\varsigma^4}\left(\left(\frac{P}{h}-\varsigma^2\right)\bm{I} +\bm{u}\otimes\bm{u}\right):\left(\bm{c}_i\otimes\bm{c}_i - \varsigma^2\bm{I}\right) + \frac{w_i^*}{4} \left(1 - \frac{P}{h\varsigma^2}\right)\right).
\end{equation}
Here $P$ is defined in Eq. \eqref{eq:P_SWE} and $w_i$ are the D2Q9 weights defined as,
\begin{equation}
    w_i = w_{c_{ix}} w_{c_{iy}},
\end{equation}
with,
\begin{equation}
   w_{c_{i\alpha}} \in \{1/6,2/3,1/6\},
\end{equation}
while $w^*_i$ can be defined through,
\begin{equation}
    w^*_i = \frac{1}{\varsigma^4}\left(c_{ix}^2-\varsigma^2\right)\left(c_{iy}^2-\varsigma^2\right).
\end{equation}
Standard second-order equilibrium amounts to a neglect of the last term in \eqref{eq:salmon_equilibrium}.
We discuss the hydrodynamic limit of this model below.
\subsection{Deviatoric components of viscous stress tensor}
A multi-scale analysis of the LBGK model \eqref{eq:coll_stream_LBGK} leads to the following momentum balance at the Navier--Stokes level, see Appendix \ref{ap:CE_LBM},
\begin{equation}
    \partial_t^{(2)} h \bm{u} + \bm{\nabla}\cdot\bm{T} = 0.
\end{equation}
Both the second-order and Salmon's equilibrium result in
the deviatoric components, $\alpha\neq\gamma$, of the non-equilibrium stress tensor as follows:
\begin{equation}
    T_{\alpha\gamma} = \frac{(1-\beta)}{2\beta}\left( \partial_\alpha h u_\gamma u_\alpha^2 + \partial_\gamma h u_\alpha u_\gamma^2 + u_\alpha \partial_\gamma P + u_\gamma \partial_\alpha P - \varsigma^2\left(\partial_\alpha h u_\gamma + \partial_\gamma h u_\alpha\right)\right).
\end{equation}
Using the definition of Eq.~\eqref{eq:beta}, and re-arranging the above viscous stress tensor, we obtain,
\begin{multline}
    T_{\alpha\gamma} = -h\nu\left(\partial_\alpha u_\gamma + \partial_\gamma u_\alpha\right)
    \\ \underbrace{+\frac{\nu}{P}\left(\partial_\alpha  u_\gamma\left( h u_\alpha^2 + (P - h\varsigma^2)\right) + \partial_\gamma  u_\alpha\left( h u_\gamma^2 + (P - h\varsigma^2)\right)\right)}_{\rm Error}.
\end{multline}
This expression reveals that the off-diagonal component of the stress tensor admits error with contributions both linear and cubic in the flow velocity.
{This error is manifest in Fig.\ \ref{Figureshear} in section \ref{sec:shear}, where it is contrasted to the result of the present lattice Boltzmann model. The latter is free of the said error.}

\subsection{Diagonal components of viscous stress tensor}
The diagonal components of the  stress tensor for the LBGK model \eqref{eq:coll_stream_LBGK} are, 
\begin{multline}\label{eq:errors_salmon}
    T_{\alpha\alpha} = -\frac{P(1-\beta)}{2\beta}\left(2\partial_\alpha u_\alpha - \bm{\nabla}\cdot\bm{u}\right) \underbrace{- \frac{P(1-\beta)}{2\beta}\left(2-\frac{h\partial P}{P\partial h}\right)\bm{\nabla}\cdot\bm{u}}_{{\rm Bulk\,viscosity\, contribution}} \\ \underbrace{+ \frac{(1-\beta)}{2 \beta} \partial_\alpha u_\alpha\left(h u_\alpha^2 + 3(P-h\varsigma^2)\right)}_{{\rm Error}},
\end{multline}
For the shallow-water equation of state \eqref{eq:P_SWE}, the bulk viscosity vanishes,
\begin{equation}
    2-\frac{h\partial P}{P\partial h} = 0,
\end{equation}
and errors remain leading to,
\begin{equation}\label{eq:bulk_2nd_pol}
    T_{\alpha\alpha} = -h\nu\left(2\partial_\alpha u_\alpha - \bm{\nabla}\cdot\bm{u}\right)  \underbrace{ + \frac{h\nu}{P} \partial_\alpha u_\alpha\left(h u_\alpha^2 + 3(P-h\varsigma^2)\right)}_{{\rm Error}}.
\end{equation}
The diagonal components of the scheme proposed in the present paper are,
\begin{equation}
    T_{\alpha\alpha} = -h\nu\left(2\partial_\alpha u_\alpha - \bm{\nabla}\cdot\bm{u}\right)  - h\eta \bm{\nabla}\cdot\bm{u},
\end{equation}
which are error-free and fully consistent with the viscous shallow water equations.
{Error of normal dissipation rate of the LBGK is manifest in Fig. \ref{normal_dissipation_rate_salmon} and is contrasted to the error-free present result in Fig.\ \ref{normal_dissipation_rate_new} of section \ref{sec:normal}.}

\section{Spectral analysis of linearized shallow water equations\label{ap:SWE_spectral}}
Here we detail the linearization process of the shallow-water equations and derive spectral dispersion and dissipation of eigen-modes. To that end, we introduce the following first-order expansion around a global equilibrium state $\left(\bar{h},\bar{\bm{u}}\right)$,
\begin{equation}
    h = \bar{h} + h',\,\, \bm{u} = \bar{\bm{u}} + \bm{u}'
\end{equation}
where $h'$ and $\bm{u}'$ are the perturbations. Introducing the expansions into the balance equation for $h$, dropping the bars for readability, and collecting linear terms,
\begin{equation}
    \partial_t h' + h\bm{\nabla}\cdot \bm{u}' + \bm{u}\cdot\bm{\nabla} h'= 0,
\end{equation}
while for the momentum balance equations,
\begin{multline}
    h\partial_t \bm{u}' + h\bm{u}\cdot\bm{\nabla}\bm{u}' + gh\bm{\nabla} h' \\ - h \nu \bm{\nabla}\cdot\left(\bm{\nabla}\bm{u}'+{\bm{\nabla}\bm{u}'}^\dagger - \frac{2}{D}\bm{\nabla}\cdot\bm{u}'\bm{I}\right) - h\eta\bm{\nabla}\cdot \left(\bm{\nabla}\cdot\bm{u}'\bm{I}\right)= 0.
\end{multline}
Next considering the perturbations to be monochromatic plane waves of the form,
\begin{equation}
    \{h',\bm{u}'\} = \{\hat{h},\hat{\bm{u}}\}\exp{[\mathrm{i}(\bm{k\cdot\bm{x}}-\omega t)]},
\end{equation}
and introducing them into the linearized equations, one obtains,
	\begin{align}
    \mathrm{i}\omega \hat{h} - \mathrm{i}h\bm{k}\cdot \hat{\bm{u}} - \mathrm{i}\bm{u}\cdot\bm{k} \hat{h} &= 0,\\
    \mathrm{i} \omega \hat{\bm{u}} - \mathrm{i} (\bm{k} \cdot\bm{u}) \hat{\bm{u}} - \mathrm{i} g \bm{k} \hat{h}+ \nu \bm{k}^2 \hat{\bm{u}} + \left(\frac{D-2}{D}\nu + \eta\right) \bm{k}(\bm{k}\cdot\hat{\bm{u}}) &= 0.
	  \end{align}
This system of equation can be written as an eigen-value problem,
\begin{equation}
    \omega \Phi = \bm{M}\Phi,
\end{equation}
with, in 2-D, $\Phi = \left[\hat{h}, \hat{u_x}, \hat{u_y}\right]^\dagger$ and,
\begin{equation}
    \bm{M} = \begin{bmatrix}
                    \bm{k}\cdot\bm{u} & h k_x & h k_y\\
                    g k_x & \bm{k}\cdot\bm{u}+\mathrm{i}(\nu \bm{k}^2 + \eta k_x^2) & \mathrm{i} \eta k_x k_y\\
                    g k_y & \mathrm{i} \eta k_x k_y & \bm{k}\cdot\bm{u}+\mathrm{i}(\nu \bm{k}^2 + \eta k_y^2)
\end{bmatrix}.
\end{equation}
Solving the eigen-value problem, $\det(\bm{M}-\omega\bm{I}){=0}$, one obtains,
	\begin{align}
    \omega_{\rm shear} &= \bm{k}\cdot\bm{u} + \mathrm{i}\nu k^2,\\
    \omega_{\rm ac+} &= \bm{k}\cdot\bm{u} +  k\sqrt{g h} + \mathrm{i}\frac{(\nu+\eta)}{2} k^2,\\
    \omega_{\rm ac-} &= \bm{k}\cdot\bm{u} -  \bm{k}\sqrt{g{h}} + \mathrm{i}\frac{(\nu+\eta)}{2} k^2,
	  \end{align}
indicating the existence of two acoustic and one shear mode.
\bibliographystyle{jfm}
\bibliography{References}
\end{document}